\documentclass[
 aps,
 prb,
 amsmath,amssymb,
 reprint,
superscriptaddress,
footinbib
]{revtex4-2}

\pdfoutput=1
\usepackage{graphicx}% Include figure files
\usepackage{dcolumn}% Align table columns on decimal point
\usepackage{bm}% bold math
\usepackage{empheq}
\usepackage[outdir=./]{epstopdf}
\usepackage{verbatim}
\usepackage{amsmath}
\usepackage{mathrsfs}
\usepackage{mathtools}
\usepackage{pdfpages}
\usepackage{pgffor}
\usepackage[caption=false]{subfig}
\newcommand\numberthis{\addtocounter{equation}{1}\tag{\theequation}}

\makeatletter
\newcommand{\Spvek}[2][r]{%
  \gdef\@VORNE{1}
  \left(\hskip-\arraycolsep%
    \begin{array}{#1}\vekSp@lten{#2}\end{array}%
  \hskip-\arraycolsep\right)}

\def\vekSp@lten#1{\xvekSp@lten#1;vekL@stLine;}
\def\vekL@stLine{vekL@stLine}
\def\xvekSp@lten#1;{\def\temp{#1}%
  \ifx\temp\vekL@stLine
  \else
    \ifnum\@VORNE=1\gdef\@VORNE{0}
    \else\@arraycr\fi%
    #1%
    \expandafter\xvekSp@lten
  \fi}
\makeatother

\makeatletter
\AtBeginDocument{\let\LS@rot\@undefined}
\makeatother

\def\supplementfilename{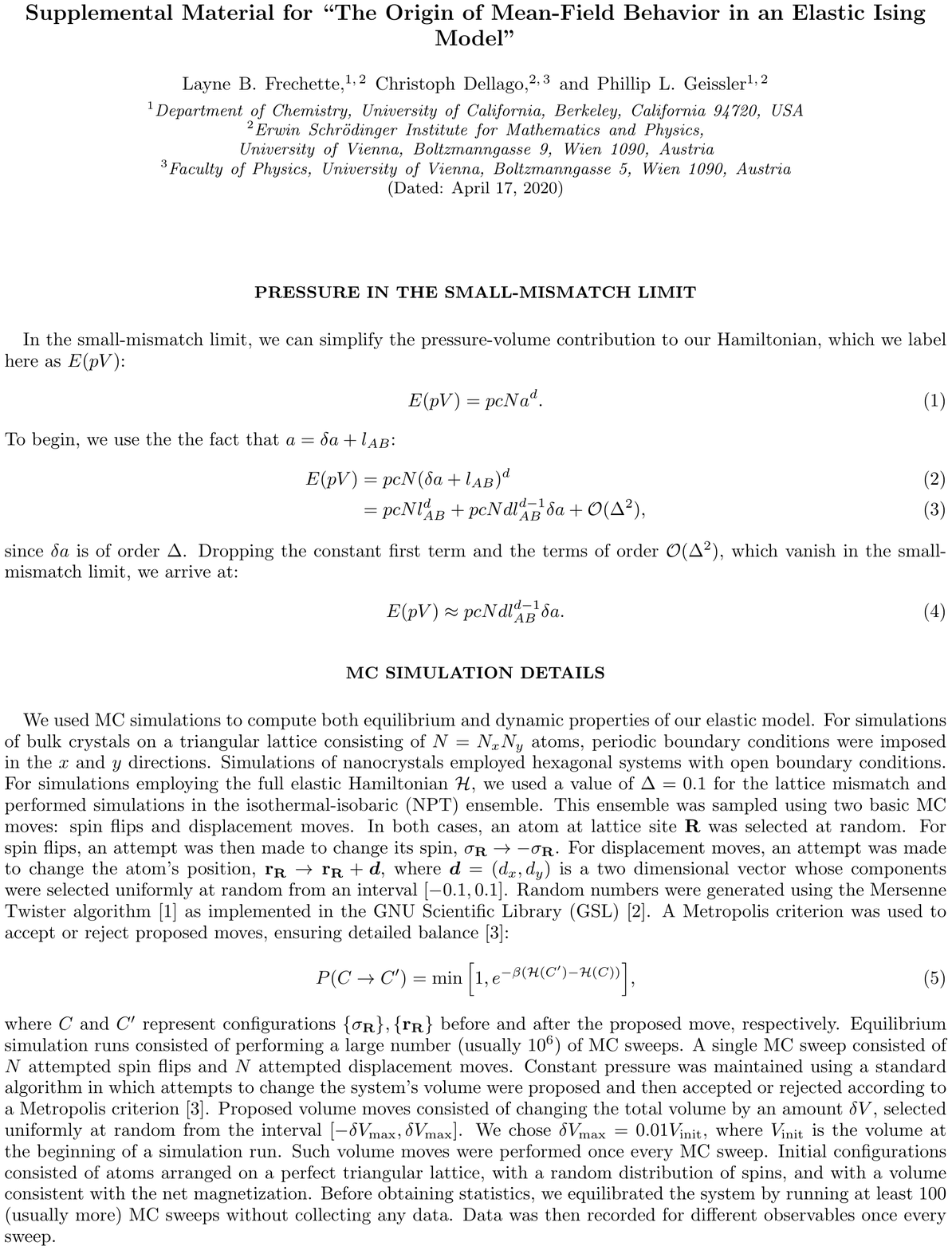}

\pdfximage{\supplementfilename}
\def\numbersupplementpages{\the\pdflastximagepages}

\newif\ifarXiv
\arXivtrue 

\begin{document}

\preprint{AIP/123-QED}

\title{The Origin of Mean-Field Behavior in an Elastic Ising Model}

\author{Layne B. Frechette}
\affiliation{Department of Chemistry, University of California, Berkeley, California 94720, USA}
\affiliation{Erwin Schr{\"o}dinger Institute for Mathematics and Physics, University of Vienna, Boltzmanngasse 9, Wien 1090, Austria}
\author{Christoph Dellago}
\email{christoph.dellago@univie.ac.at }
\affiliation{Erwin Schr{\"o}dinger Institute for Mathematics and Physics, University of Vienna, Boltzmanngasse 9, Wien 1090, Austria}
\affiliation{Faculty of Physics, University of Vienna, Boltzmanngasse 5, Wien 1090, Austria}
\author{Phillip L. Geissler}
\email{geissler@berkeley.edu}
\affiliation{Department of Chemistry, University of California, Berkeley, California 94720, USA}
\affiliation{Erwin Schr{\"o}dinger Institute for Mathematics and Physics, University of Vienna, Boltzmanngasse 9, Wien 1090, Austria}

\date{\today}

\begin{abstract}
	
Simple elastic models of spin-crossover compounds are known
empirically to exhibit classical critical behavior. We demonstrate how
the long-ranged interactions responsible for this behavior arise
naturally upon integrating out mechanical fluctuations of such a
model. A mean field theory applied to the resulting effective
Hamiltonian quantitatively accounts for both thermodynamics and
kinetics observed in computer simulations, including a barrier
to magnetization reversal that grows extensively with system size. For
nanocrystals, which break translational symmetry, a straightforward
extension of mean field theory yields similarly accurate results.

\end{abstract}

%\keywords{Statistical mechanics, elasticity, lattice models, phase separation}%Use showkeys class option if keyword
\maketitle

\section{Introduction}

The impact of spin-lattice interactions on materials’ phase behavior has long been a topic of interest in condensed matter physics and materials science \cite{Rice1954,Domb1956,Baker1970,Oitmaa1975}. Microscopic coupling between spin and geometry in an extended material can endow it with intriguing and useful properties, such as susceptibility of crystal structure to light or pressure \cite{Letard1999,Hauser1999,Real2005,GUTLICH2005,Konishi2008}. Elastic Ising models provide a minimal representation of such materials. In a simple variant, the atoms of a crystal lattice interact with their neighbors via Hookean springs. The natural length of these springs is determined by the participating atoms' internal ``spin'' (which could represent either a literal spin state or a chemical identity.) This type of model has been employed in studies of lattice-mismatched semiconductor alloys \cite{Dunweg1993, Vandeworp1997} and spin-crossover compounds \cite{Miyashita2008, Miyashita2009}. Despite its substantial history, one of the most basic aspects of this model's behavior remains unresolved. The aforementioned studies employed Monte Carlo (MC) simulations to demonstrate that elastic Ising models can exhibit demixing transitions governed by mean-field critical exponents. However, the microscopic origin of this behavior has not been explicitly identified, nor has a quantitative framework for predicting its consequences been developed.

Here, we present a thorough explanation for the origin of this mean-field behavior. Drawing from our recent work on a similar elastic Ising model \cite{Frechette2019}, we show how the coupling of mechanical fluctuations to spins engenders effective inter-atomic interactions with infinite spatial extent. These give rise to the observed mean-field critical behavior. With an explicit form for the interactions in hand, we develop a straightforward mean field theory (MFT) which accurately predicts the free energy as a function of magnetization as well as the critical temperature for spontaneous symmetry breaking. MFT yields similarly faithful predictions for relaxation dynamics of the magnetization in the presence of an external field. Finally, we extend our theory to describe spatially heterogeneous systems such as nanocrystals. Our results provide a theoretical basis not only for interpreting the results of a number of previous computer simulation studies, but also for the design of switchable elastic materials.

\section{Elastic Ising Model and Effective Interactions}
We consider a collection of $N$ atoms at positions $\mathbf{r}_{\mathbf{R}}=\mathbf{R}+\mathbf{u}_{\mathbf{R}}$. The quantity $\mathbf{R}$ denotes a site on a $d$-dimensional crystal lattice characterized by unit bond vectors $\bm{\hat{\alpha}}$, and $\mathbf{u}_{\mathbf{R}}$ is the displacement of an atom
from its ideal lattice site. Spin variables $\sigma_{\mathbf{R}}=\pm 1$ determine the natural bond length between neighboring atoms:
\begin{equation}
l(\sigma_{\mathbf{R}},\sigma_{\mathbf{R}+a\bm{\hat{\alpha}}}) = \begin{cases} l_{AA},\,\,\text{for}\,\, \sigma_{\mathbf{R}}=\sigma_{\mathbf{R}+a\bm{\hat{\alpha}}} = 1\\
l_{AB},\,\,\text{for}\,\,\sigma_{\mathbf{R}} \neq \sigma_{\mathbf{R}+a\bm{\hat{\alpha}}}\\
l_{BB},\,\,\text{for}\,\,\sigma_{\mathbf{R}} = \sigma_{\mathbf{R}+a\bm{\hat{\alpha}}} = -1,
\end{cases}
\end{equation} 
where $a$ is the fluctuating lattice parameter, $l_{BB}<l_{AA}$, and
$l_{AB}=(l_{AA}+l_{BB})/2$. We choose the lattice mismatch $\Delta =
(l_{AA}-l_{BB})/2$ to be our basic unit of length. An external pressure $p$ couples directly to the volume $cNa^d$, where $c$ is a geometry-dependent constant of $\mathcal{O}(1)$. The Hamiltonian governing the system is quadratic in deviations of bond lengths $|\mathbf{r}_{\mathbf{R}+a\hat{\bm{\alpha}}}-\mathbf{r}_{\mathbf{R}}|$ from their preferred $\sigma$-dependent values:
\begin{equation}
\mathcal{H}=\frac{K}{4}\sum_{\mathbf{R},\bm{\hat{\alpha}}}\left[|a\bm{\hat{\alpha}}
+\mathbf{u}_{\mathbf{R}+a\bm{\hat{\alpha}}}-\mathbf{u}_{\mathbf{R}}|-l(\sigma_{\mathbf{R}}
,\sigma_{\mathbf{R}+a\bm{\hat{\alpha}}})\right]^2 +pcNa^d.
\label{eq:hamil}
\end{equation}
The spring constant $K>0$ determines the elastic energy scale $\epsilon=K\Delta^2/8$. We express all quantities henceforth in units of $\Delta$ and $\epsilon$. Eq. \ref{eq:hamil} manifestly couples spin and displacement variables. We will show how the effect of fluctuations in the displacements can be captured by an effective energy function
$\mathcal{H}_{\text{eff}}$ of the spin variables:
\begin{align}
\mathcal{H}_{\text{eff}}[\{\sigma_{\mathbf{R}}\}]&=\mathcal{H}^{\text{SR}} + \mathcal{H}^{\text{LR}} - h\sum_{\mathbf{R}}\sigma_{\mathbf{R}}\label{eq:eff_hamil_form}\\
\mathcal{H}^{\text{SR}}&=\frac{1}{2}\sum_{\mathbf{R},\mathbf{R}'}\sigma_{\mathbf{R}}V_{\mathbf{R}-\mathbf{R}'}^{\text{SR}}\sigma_{\mathbf{R}'}\\
\mathcal{H}^{\text{LR}}&=\frac{1}{2N}V^{\text{LR}}\left(\sum_{\mathbf{R}}\sigma_{\mathbf{R}}\right)^2,
\end{align}
where ``SR'' and ``LR'' stand for ``short-ranged'' and ``long-ranged,'' respectively. 
$V_{\mathbf{R}}^{\text{SR}}$ is an effective interaction potential that decays
steadily with distance $|\mathbf{R}|$, and $V^{\text{LR}}$ is a constant that sets the strength of long-range coupling.
This form of spin interactions guarantees mean field critical behavior,
as will be discussed below.

We first simplify Eq. \ref{eq:hamil} by noting that if $\Delta$ is small, $\mathcal{H}$ can be written approximately as (see \cite{Frechette2019} and the Supplemental Material \cite{SM}):
\begin{multline}
\mathcal{H} \approx 2\sum_{\mathbf{R},\hat{\bm{\alpha}}}\left(\hat{\bm{\alpha}}\cdot \left(\mathbf{u}_{\mathbf{R}+a\hat{\bm{\alpha}}}
-\mathbf{u}_{\mathbf{R}}\right) - \frac{1}{2}( \delta\sigma_{\mathbf{R}} +
\delta\sigma_{\mathbf{R}+a\hat{\bm{\alpha}}})\right.\\\left.-(\tilde{\sigma}_0/N-\delta a)\right.\biggr)^2 - Nh\delta a,
\label{eq:small_disp}
\end{multline}
where $h=-pcdl_{AB}^{d-1}$ is a dimensionless pressure
and $\delta a = a-l_{AB}$. We have partitioned the spin variables into two components, namely the net magnetization $\tilde{\sigma}_0=\sum_{\mathbf{R}}\sigma_{\mathbf{R}}$ and the local deviation $\delta \sigma_{\mathbf{R}}=\sigma_{\mathbf{R}}-\tilde{\sigma}_0/N$. Using $\sum_{\mathbf{R}}\mathbf{u}_{\mathbf{R}}=0$, we expand Eq. \ref{eq:small_disp}:
\begin{equation}
\mathcal{H} = \Delta \mathcal{H}(\{\mathbf{u}_{\mathbf{R}}\}), \{\delta \sigma_{\mathbf{R}}\}) + 2(\tilde{\sigma}_0/N-\delta a)^2NZ - Nh\delta a,
\end{equation}
where $Z$ is the coordination number of the lattice and
\begin{equation}
\Delta \mathcal{H} = 2\sum_{\mathbf{R},\hat{\bm{\alpha}}}\left(\hat{\bm{\alpha}}\cdot \left(\mathbf{u}_{\mathbf{R}+a\hat{\bm{\alpha}}}
-\mathbf{u}_{\mathbf{R}}\right) - \frac{1}{2}( \delta\sigma_{\mathbf{R}} +
\delta\sigma_{\mathbf{R}+a\hat{\bm{\alpha}}})\right)^2.
\end{equation}
Gaussian fluctuations in the lattice parameter $\delta a$ evidently couple solely to $\tilde{\sigma}_0$. Working in an ensemble with fixed $N$, $p$, and inverse temperature $\beta=1/k_BT$, where $k_B$ is Boltzmann's constant, we integrate out these fluctuations:
\begin{align}
\bar{\mathcal{H}}&= -\beta^{-1}\log\left(\int d(\delta a)\exp{(-\beta \mathcal{H})}\right)\\
&= \Delta \mathcal{H} - h\tilde{\sigma}_0+\text{const}.\label{eq:hamil_bar}
\end{align}
    We see that $h$
    simply plays the role of an effective field acting on $\tilde{\sigma}_0$, and so spin coupling is contained entirely in $\Delta \mathcal{H}$. We interrogate this coupling by further integrating out Gaussian fluctuations in the displacement field (dropping the unimportant constant term in Eq. \ref{eq:hamil_bar}):
\begin{equation}
\mathcal{H}_{\text{eff}}=-\beta^{-1}\log{\left(\int \prod_{\mathbf{R}} d\mathbf{u}_{\mathbf{R}} \exp({-\beta \Delta \mathcal{H}})\right)}-h\tilde{\sigma}_0.
\label{eq:effhamil_int}
\end{equation}
If we assume that our system is subject to periodic boundary conditions, then the required integrals are most easily performed in Fourier space. This yields (see \cite{Frechette2019}):
\begin{equation}
\mathcal{H}_{\text{eff}}[\{\sigma_{\mathbf{R}}\}] = \frac{1}{2N}\sum_{\mathbf{q}} \tilde{V}_{\mathbf{q}}|\tilde{\sigma}_{\mathbf{q}}|^2 - h \tilde{\sigma}_0,
\label{eq:effhamil}
\end{equation}
where $\tilde{f}_{\mathbf{q}}$ denotes the Fourier transform of a generic function $f_{\mathbf{R}}$ \footnote{Note that we have dropped the $\delta$ in front of $\tilde{\sigma}_{\mathbf{q}}$. That is because $\delta \tilde{\sigma}_{\mathbf{q}}=\tilde{\sigma}_{\mathbf{q}}-\delta_{\mathbf{q},0}\tilde{\sigma}_{0}$, but $\tilde{V}_0=0$, so $\tilde{\sigma}_0$ simply does not contribute to the sum.}.
The explicit form of the effective potential $\tilde{V}_{\mathbf{q}}$ for the triangular lattice is given by \cite{Frechette2019}:
\begin{equation}
\tilde{V}_{\mathbf{q}} = \begin{cases}
\frac{4\left(2\cos{\frac{q_xa}{2}}\cos{\frac{\sqrt{3}q_ya}{2}}+\cos{q_xa}-3\right)^2}{\left(\cos{q_xa}-2\right)\left(4\cos{\frac{q_xa}{2}}\cos{\frac{\sqrt{3}q_ya}{2}}-3\right)+\cos{\sqrt{3}q_ya}}, & \mathbf{q}\neq 0\\
0, & \mathbf{q}=0,
\end{cases}
\label{eq:tri_lat_pot}
\end{equation}
where $q_x$ and $q_y$ are the Cartesian components of $\mathbf{q}$.

\begin{figure}
	\centering
	\includegraphics[width=\linewidth]{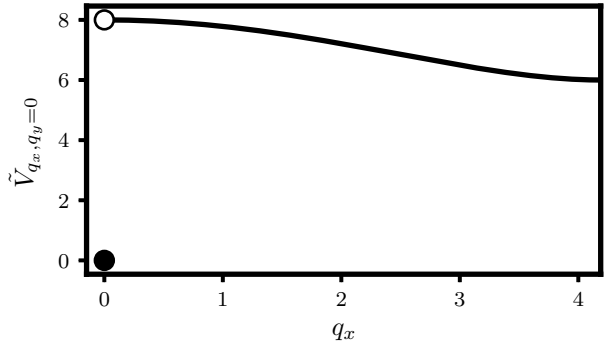}
	\caption{Fourier-space effective potential for the triangular lattice, Eq. \ref{eq:tri_lat_pot}. Note that $\tilde{V}_\mathbf{q}$ is smooth everywhere except $\mathbf{q}=0$, since $\tilde{V}_0=0$, but $\lim_{\mathbf{q}\rightarrow 0}\tilde{V}_{\mathbf{q}}=8$.}
	\label{fig:veff}
\end{figure}

The existence of a long-ranged coupling is not immediately evident
from this analysis,
since the longest-wavelength component of the potential ($\tilde{V}_0$) is zero. However, the limit of the potential as $\mathbf{q}\rightarrow 0$ is not approached smoothly (see Fig. \ref{fig:veff}) a required condition for short-ranged interactions \cite{Dantchev2001,Stein2003}. Observe that a simple modification of $\tilde{V}_{\mathbf{q}}$ \textit{does} vanish smoothly as $\mathbf{q}\rightarrow 0$:
\begin{equation}
\tilde{V}_{\mathbf{q}}^{\text{SR}}=\tilde{V}_{\mathbf{q}}-(1-\delta_{\mathbf{q},0})\lim_{\mathbf{q}'\rightarrow 0 }\tilde{V}_{\mathbf{q}'}.
\end{equation} 
Its inverse transform $V_{\mathbf{R}}^{\text{SR}}$ is therefore a well-defined short-ranged interaction \footnote{The function $V_{\mathbf{R}}^{\text{SR}}$ is generally anisotropic; for the triangular lattice, its slowest decay is $1/|\mathbf{R}|^4$ along (certain linear combinations of) triangular lattice basis vectors \cite{Frechette2019}.}. The remainder of $\tilde{V}_{\mathbf{q}}$ is:
\begin{align}
\tilde{V}_{\mathbf{q}}^{\text{LR}}&=\tilde{V}_{\mathbf{q}}-\tilde{V}_{\mathbf{q}}^{\text{SR}}\\
&= \text{const.} - \delta_{\mathbf{q},0}\lim_{\mathbf{q}'\rightarrow 0}\tilde{V}_{\mathbf{q}'},
\end{align}
where the constant term simply generates an irrelevant self-interaction, which we drop. Plugging this back into the sum in Eq. \ref{eq:effhamil} and writing all quantities in terms of real-space sums gives us the promised form Eq. \ref{eq:eff_hamil_form}, with:
\begin{equation}
V^{\text{LR}}=-\lim_{\mathbf{q}\rightarrow 0}\tilde{V}_{\mathbf{q}}.
\end{equation}
The limit is given explicitly by:
\begin{align}
\lim_{\mathbf{q}\rightarrow 0}\tilde{V}_{\mathbf{q}}&=
2Z-4\mathbf{a}(\hat{\mathbf{q}})\cdot \mathbf{A}^{-1}(\hat{\mathbf{q}})\cdot \mathbf{a}(\hat{\mathbf{q}})\\
\mathbf{a}(\hat{\mathbf{q}})&=\sum_{\hat{\bm{\alpha}}}(\hat{\mathbf{q}}\cdot\hat{\bm{\alpha}})\hat{\bm{\alpha}}\\
\mathbf{A}(\hat{\mathbf{q}})&= \sum_{\hat{\bm{\alpha}}}(\hat{\mathbf{q}}\cdot \hat{\bm{\alpha}})^2\hat{\bm{\alpha}}\hat{\bm{\alpha}},
\end{align}
where $\hat{\mathbf{q}}$ is an arbitrary unit vector.
For the triangular lattice, this simplifies to $\lim_{\mathbf{q}\rightarrow 0}\tilde{V}_{\mathbf{q}}=8$.

\section{Mean Field Theory}

That long-ranged interactions are operative in spin-crossover compounds has been suggested by several authors \cite{Willenbacher1988,Spiering1989,Kohler1990,Boukheddaden2000,Boukheddaden2000a,Fourati2018}. Miyashita et al. \cite{Miyashita2008} conjectured that the long-ranged interactions responsible for mean-field behavior in their model had the same $1/|\mathbf{R}-\mathbf{R}'|^3$
decay
as that between point defects in three-dimensional continuum elastic media.
We have demonstrated that,
instead, an infinitely long-ranged interaction arises from a discontinuity in the spectrum $\tilde{V}_\mathbf{q}$. This nonanalytic feature originates physically in a mismatch between the elastic energy associated with $\mathbf{q}=0$ and small (but nonzero) wavevector variations in the magnetization. Schulz et al. \cite{Schulz2005} argued that precisely those long-wavelength elastic modes ought to be responsible for the mean-field behavior of elastic models of binary alloys. 

A model which consists solely of interactions of the form $\mathcal{H}^{\text{LR}}$ is described exactly by MFT \cite{Kac1963,Cannas2000,Vollmayr-Lee2001,Mori2010}. Its mean-field critical exponents are robust to the addition of arbitrary short-ranged interactions \cite{Capel1979,Nakada2011} (a fact which we confirmed numerically for several different lattice structures; see \cite{SM} for details). There is no such guarantee for non-universal quantities such as the critical temperature $T_c$, but if the magnitude of $V_{\mathbf{R}}^{\text{SR}}$ is small, then MFT may still predict their values with reasonable accuracy. We obtained such predictions using standard techniques of MFT \cite{Chandler1987}, which yield a self-consistent equation for the net magnetization per atom $m=\tilde{\sigma}_{0}/N$:
\begin{equation}
m = \tanh{(2\beta \bar{V} m+h)},
\label{eq:sc_field}
\end{equation}
as well as a simple expression for the free energy $F(m)$:
\begin{align}
F_{\text{MF}}(m)&=E_{\text{MF}}(m)-TS_{\text{MF}}(m)\label{eq:free_energy}\\
E_{\text{MF}} &= -N\bar{V}m^2-N hm \label{eq:mft_energy}\\
S_{\text{MF}}/k_B &= \ln \binom{N}{N\frac{1+m}{2}}\\
&\approx
-N\left[\frac{1-m}{2}\log{\frac{1-m}{2}} + \frac{1+m}{2}\log{\frac{1+m}{2}}\right],
\end{align}
 where $\bar{V}=-\sum_{\mathbf{R}\neq 0}V_{\mathbf{R}}/2$ characterizes both long- and short-ranged
contributions to the mean field.  The second expression for
$S_{\text{MF}}$, obtained from Stirling's approximation for large $N$,
will be used in mean-field calculations that do not specify system
size.
When $h=0$, Eq. \ref{eq:sc_field} implies a critical temperature $T_c = 2\bar{V}$ for spontaneous symmetry breaking.

\begin{figure}
	\centering
	\includegraphics{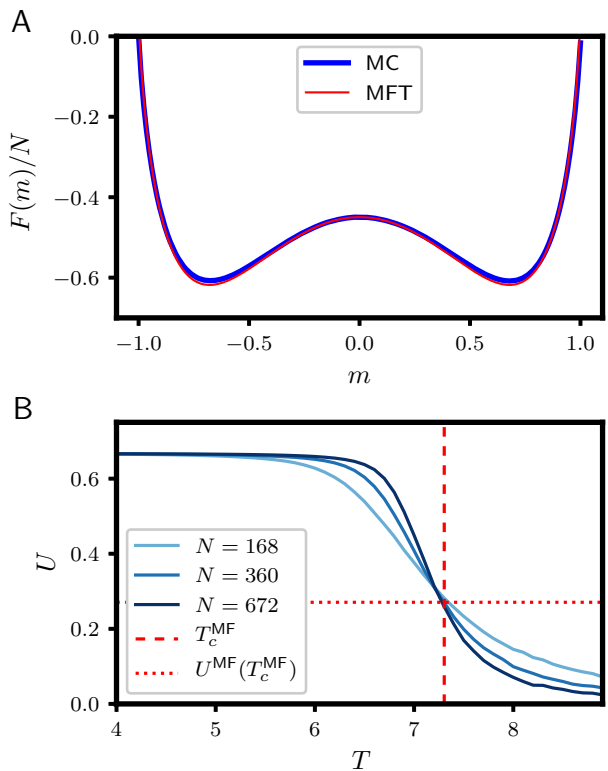}
	\caption{Comparison between MC and MFT results for the
          free energy as a function of magnetization and for the critical
          temperature. \textbf{A}: Free energy profile for a
          periodic triangular lattice with $N=168$ at $T=6$, $p=0$. The MC curve was
          computed with umbrella sampling for the model of
          Eq. \ref{eq:hamil} (see \cite{SM} for details), while the
          MFT curve was obtained from
          Eq. \ref{eq:free_energy}. \textbf{B}: MFT estimates for the
          triangular lattice critical temperature
          $T_c^{\text{MF}}=2\bar{V}\approx 7.31$ and the corresponding
          value of the Binder cumulant
          $U^{\text{MF}}(T_c^{\text{MF}})$ \cite{Miyashita2008}
          closely predict the intersection point of MC Binder
          cumulants $U$ for different system sizes. Specifically, MC
          indicates that $T_c^{\text{MC}}\approx 7.2$, so the MFT
          result is accurate within $\approx 2$\%. MC results for the
          Binder cumulants were computed by sampling $10^6$
          configurations at each temperature. These configurations
          were generated using the effective energy function of
          Eq. \ref{eq:effhamil}
          rather than Eq. \ref{eq:hamil}, in order to avoid statistical
          errors associated with insufficient sampling of mechanical
          fluctuations.}
	\label{fig:free}
\end{figure}

To test the accuracy of MFT, we performed MC simulations of the elastic Hamiltonian Eq. \ref{eq:hamil} on a periodic triangular lattice. Specifically, we repeatedly proposed changes in $a$, $\{\mathbf{u}_\mathbf{R}\}$, and $\{\sigma_{\mathbf{R}}\}$ and accepted these changes with probabilities designed to satisfy detailed balance (see \cite{SM} for details.) Some simulations (described later) were performed using the effective Hamiltonian Eq. \ref{eq:effhamil}; for these simulations, only changes in $\{\sigma_{\mathbf{R}}\}$ were necessary. We employed umbrella sampling \cite{Torrie1977} combined with histogram reweighting \cite{Kumar1992} to compute free energies. In addition, we located $T_c$ from the intersection of Binder cumulants computed at different system sizes \cite{Binder1981a}. The results agree quantitatively with our mean-field predictions, as shown in Fig. \ref{fig:free}. We found similarly excellent agreement between MC and MFT for several different lattice structures in both two and three dimensions \cite{SM}.

\section{Dynamics: Quenching and Hysteresis}

As a more stringent test of MFT, we consider dynamics of our elastic Ising model. Free energy profiles like that in Fig. \ref{fig:free} are suggestive of time-dependent response that would follow a sudden change in external control parameters. But this relaxation advances in the
high-dimensional space of spin configurations, through sequential flips of spins that are correlated in space and in time. Resolving few of these details, MFT asserts that thermodynamic driving forces determine time evolution in a simple way. Its success in a dynamical context would provide powerful tools to predict and understand nonequilibrium response.

The model energy function in Eq. \ref{eq:hamil} constrains microscopic rules for time evolution but does not specify them uniquely. To craft a dynamical model we must additionally assign rates to microscopic transitions which are consistent with Boltzmann statistics. As a simplification, we take relaxation of the
displacement variables $\mathbf{u}_\mathbf{R}$ to be much faster than that of spin variables. This rapid mechanical equilibration allows us to consider time variations of the spin field $\sigma_\mathbf{R}$ alone, biased by an effective Hamiltonian. In the small-mismatch limit this effective energy is given by Eq. \ref{eq:eff_hamil_form}. We adopt transition rates $\pi(\sigma_\mathbf{R}\rightarrow\sigma_\mathbf{R}')$ among spin configurations proportional to their Metropolis Monte Carlo acceptance probabilities,
$\pi(\sigma_\mathbf{R}\rightarrow
\sigma_\mathbf{R}') = \tau^{-1} \min[1,\exp{[-\beta \Delta \mathcal{H}_{\rm eff}]}]$,
where $\tau$ is an arbitrary reference time scale.

\begin{figure}
	\centering
	\includegraphics{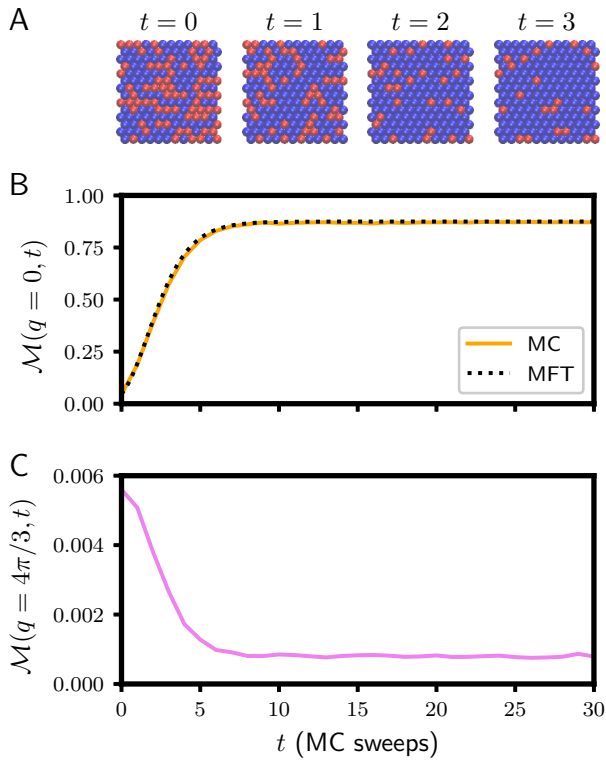}
	\caption{Magnetization dynamics after a quench from temperature $T_0=8$ to $T=4$ on the triangular lattice. \textbf{A}: Representative configurations from a single quench trajectory with $N=168$. The time $t$ following the quench is measured in MC sweeps.
          \textbf{B}: Time evolution of the $q=|\mathbf{q}|=0$ (longest-wavelength) Fourier component of the spin structure factor $\mathcal{M}(\mathbf{q},t)$ following the quench. This mode grows rapidly at short times and saturates at the equilibrium value of $m^2$. Solving the mean-field master equation, Eq. \ref{eq:master_eq}, for a system size $N=168$ and
          initial condition $P(m,0)=\exp(-
          F_{\text{MF}}(m;T=T_0)/k_BT_0)$ yields a prediction
          $\mathcal{M}(0,t)=\sum_{m=-1}^1m^2P(m,t)$ (labeled MFT in
          the plot) which closely agrees with the MC
          result. \textbf{C}: Time evolution of a short-wavelength
          Fourier component of $\mathcal{M}$ with $q=4\pi/3$
          (corresponding to a corner of the first Brillouin zone of
          the triangular lattice) computed from MC simulations. This
          mode decays rapidly, consistent with the apparent lack of
          short-wavelength structure in the configurations. MC curves
          in both \textbf{B} and \textbf{C} were obtained by averaging
          over $10^3$ independent trajectories initialized from
          equilibrium configurations sampled at $T_0=8$, and
          propagated with Metropolis spin-flip dynamics at $T=4$. All
          MC simulations here were performed using
          $\mathcal{H}_{\text{eff}}$, Eq. \ref{eq:effhamil}.}
	\label{fig:quench}
\end{figure}

The ordering dynamics that follow a rapid quench from $T>T_c$ to
$T<T_c$ are strongly influenced by the long-range component of $\mathcal{H}_{\rm eff}$. Models with exclusively short-ranged interactions, such as described by $\mathcal{H}^{\rm SR}$, develop finite-wavelength instabilities upon such quenching \cite{Cahn1965}. These Ising-like instabilities are visually manifest in the coarsening of spin domains en route to a state of broken symmetry. By contrast, a model with exclusively infinite-range interactions, such as described by $\mathcal{H}^{\rm LR}$, lacks
finite-wavelength
spatial correlations entirely and therefore does not exhibit a slowly growing length scale upon quenching. In dynamical simulations of our elastic Ising model, we observe no distinct domain growth upon quenching from $T=8>T_c$ to $T=4<T_c$, consistent with the observations of Miyashita et al. \cite{Miyashita2008}.
Indeed, our measurements of the time-dependent spin structure factor
$\mathcal{M}(\mathbf{q},t)=\langle|\tilde{\sigma}(\mathbf{q},t)|^2\rangle/N^2$ (where $\langle \cdots \rangle$ denotes an ensemble average) show that only the $\mathbf{q}=0$ mode becomes unstable upon quenching
(Fig. \ref{fig:quench}). This can be understood in detail as a
consequence of the energy gap between the $\mathbf{q}=0$ mode and the finite-wavelength modes depicted in Fig. \ref{fig:veff}. The lack of participation of the finite-wavelength modes in the quench dynamics
suggests that a mean-field framework -- in which the only dynamical
quantity is the net magnetization -- should provide a sensible description of our model's dynamical features. Indeed, a mean-field master equation, to be described below, predicts the dynamics of $\mathcal{M}(\mathbf{q}=0,t)$ very accurately (see Fig. \ref{fig:quench}B.)

\begin{figure}
	\centering
	\includegraphics{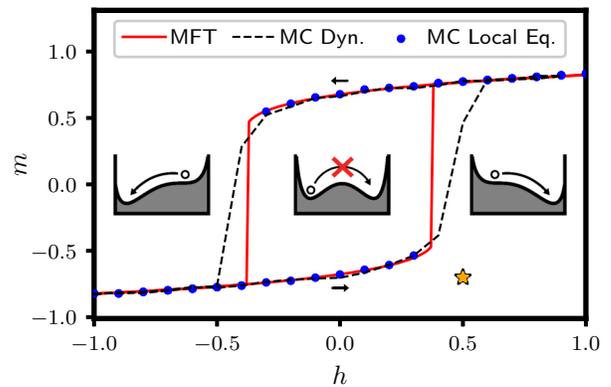}
	\caption{Hysteresis loop at $T=6$ from MFT and MC. The MFT curve was obtained by numerical solution of Eq. \ref{eq:sc_field}. Solutions to this equation which are also local free energy minima (of which there are at least one and at most two) comprise the mean-field hysteresis loop.  MC Dyn. (dynamic) results were obtained by sweeping the field from $h=-1.0$ to $h=0.9$ and back again (direction indicated by the black arrows) for a simulated system with $N=2688$ using $\mathcal{H}_{\text{eff}}$ (Eq. \ref{eq:effhamil}). For each field value, there were 10 MC sweeps of equilibration and 10 MC sweeps of production. MC Eq. (equilibrium) results were obtained by locating the local minima of the free energy as a function of magnetization (computed with umbrella sampling using $\mathcal{H}$ (Eq. \ref{eq:hamil}.)) for different values of $h$ and a system size of $N=168$. Inset schematics illustrate the fact that, in the thermodynamic limit, barrier crossing does not occur; the system can only escape from a metastable well once it has reached the limits of metastability. The yellow star indicates the values of $m$ and $h$ used as a starting point for dynamics in Fig. \ref{fig:avg_mag}.}
	\label{fig:hysteresis}
\end{figure}

In addition to changes in temperature, one can probe a system's
response to a cyclically varying parameter that crosses and recrosses
a phase boundary. In the resulting loop, the distinctness of forward
and backward branches reports on the system's ``memory'' owing to a
slow degree of freedom (the net magnetization, in our case.) If such
an experiment were performed sufficiently slowly, one would normally
expect differences between the two branches to vanish.  For our model,
hysteresis instead appears to persist for arbitrarily slow
cycling. Normally, the free energy barrier for nucleating a stable
phase is subextensive in scale, since the thermodynamic cost is
interfacial in nature. For our model, finite-size scaling of MC
simulation results indicate that the barrier separating minima in
$F(m)$ instead scales linearly with system size $N$.  This feature is inherent
to MFT, which presumes spatial heterogeneity and thus a lack of
interfaces. Thermal fluctuations are insufficient to overcome such an
extensive barrier in the thermodynamic limit, and so the system will persist indefinitely in the state in which it was initialized.  The mean-field nature of the barrier in
$F(m)$ is reflected in Fig. \ref{fig:hysteresis}, which shows
excellent agreement between hysteresis loops computed from MC and the
corresponding prediction of MFT.

At the ends of the hysteresis loop,
$F(m)$ is no longer bistable,
and a system
initialized at the location of the formerly-metastable well $m_\text{i}$
can relax
to the single stable well at $m_{\text{f}}$
without crossing a barrier.
Within MFT, this dynamics can be regarded as a
random walk of the magnetization with step length $\Delta m=2/N$ taken
in discrete time steps $\Delta t$ on the mean-field energy
surface. The probability distribution $P(m,t)$ for the magnetization
at time $t$ is governed by a master equation \footnote{Mean-field
  dynamics can also be accessed by directly simulating the MC dynamics
  of a mean-field Hamiltonian; see \cite{SM} for
  details.}:
\begin{align*}
P(m,t) &= P(m-\Delta m, t-\Delta t)\Pi_+(m-\Delta m)\\ 
&+ P(m+\Delta m, t-\Delta t)\Pi_-(m+\Delta m)\\
&+ P(m,t-\Delta t)\left(1-(\Pi_+(m)+\Pi_-(m))\right).\numberthis \label{eq:master_eq}
\end{align*}
with
transition rates
\begin{equation}
\Pi_{\pm}(m) = \frac{1\mp m}{2}\min{\left[1,e^{-\beta(E_{\text{MF}}(m\pm \Delta m)-E_{\text{MF}}(m))}\right]}, \label{eq:transition_prob}
\end{equation}
for incrementally increasing (decreasing) $m$.
The factor $(1\mp m)/2$ accounts for
the number of available down (up) spins at magnetization $m$, which imposes
an entropic bias at the mean-field level.
These
rates satisfy detailed balance
with respect to the probability distribution $e^{-\beta F_{\text{MF}}(m)}$.
The relaxation process of interest is defined by
boundary conditions:
\begin{align}
P(m,0)&=\begin{cases}
1, & m=m_{\text{i}}\\
0, & \text{otherwise}
\end{cases}\\
P(m_{\text{f}},t)&=0,
\end{align}
ensuring
that the system always begins at $m=m_{\text{i}}$, and the magnetization can never exceed $m=m_{\text{f}}$. A different set of boundary conditions was used to compute $\mathcal{M}(0,t)$ for Fig. \ref{fig:quench} (details in the corresponding caption.)

\begin{figure}
	\centering
	\includegraphics{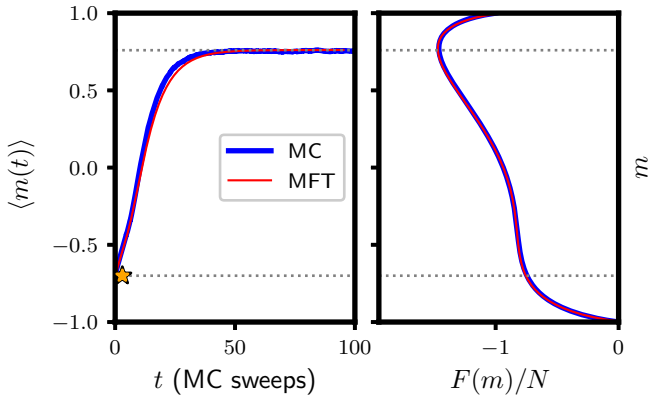}
	\caption{Average magnetization versus time for phase change
          dynamics at the end of the hysteresis loop (left) with
          $T=6$, and the corresponding free energy profile (right)
          with $T=6$, $h=0.5$. Gray dashed lines indicate the
          positions of the formerly-metastable well and the single
          stable well. The yellow star indicates the initial state,
          marked with the same symbol in
          Fig. \ref{fig:hysteresis}. The MFT result for $\langle m(t)
          \rangle$ was calculated via Eq. \ref{eq:avg_mag} for a
          system size $N=168$, and the mean-field free energy is given
          by Eq. \ref{eq:free_energy}. The MC result for $\langle m(t)
          \rangle$ was computed by averaging over $10^4$ independent
          trajectories of a system with $N=168$. These trajectories
          were propagated by Metropolis MC according to 
          the effective energy $\mathcal{H}_{\text{eff}}$
          (Eq. \ref{eq:effhamil}). Their initial configurations were
          sampled from an equilibrium trajectory whose magnetization
          $m=-0.7$ was fixed by performing Kawasaki dynamics
          \cite{Kawasaki1966}.  
          The MC result for the
          free energy was computed via umbrella sampling of a system
          with $N=168$ using $\mathcal{H}$ (Eq. \ref{eq:hamil}.)}
	\label{fig:avg_mag}
\end{figure}

\begin{figure}
	\centering
	\includegraphics{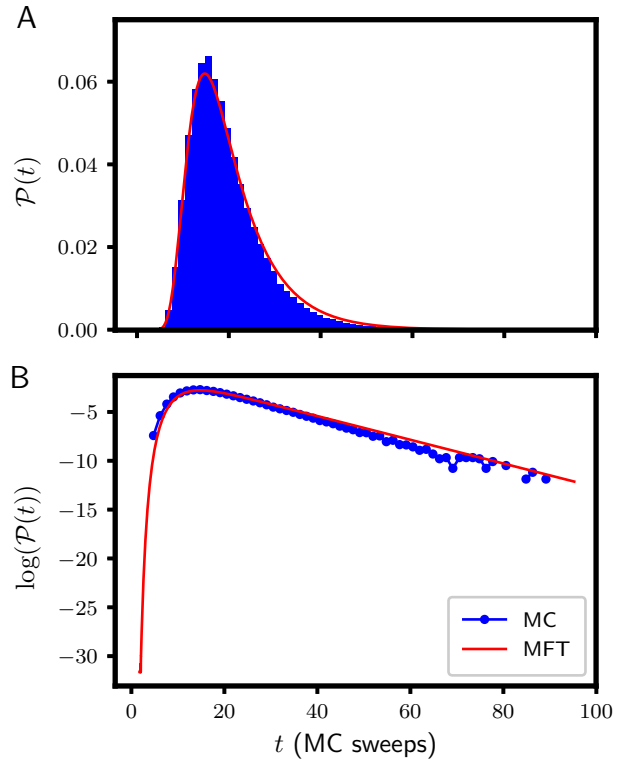}
	\caption{Probability distribution of the first passage
          time for phase change at the end of the hysteresis loop
          with $T=6$,
          plotted on linear
          (\textbf{A}, top) and logarithmic (\textbf{B}, bottom)
          scales.  For each MC trajectory, the first passage time was
          defined as the number of MC steps taken en route 
          from the formerly-metastsable initial state ($m=-0.7$) to
          the bottom of the stable well
          ($m=0.76$).
          Results are shown for a system of size
          $N=168$. MFT predictions were computed via Eq. \ref{eq:fpt}
          for the same system size. The long-time exponential tail is
          characteristic of diffusion on a bounded interval; its
          corresponding decay rate is set by the least negative eigenvalue
          of $\bm{\Omega}$ \cite{Redner2001}. The entire eigenvalue spectrum
          of $\bm{\Omega}$ is sensitive to changes in $N$, highlighting
          a size dependence of first passage time statistics that
          appears to be well captured by MFT.}
	\label{fig:fpt_dist}
\end{figure}

Defining the column vector:
 \begin{equation}
\mathbf{P}(t) 
= \left(P(-1,t),P(-1+\Delta m, t),\dots P(1,t)\right)^T,
\end{equation}
where the superscript $T$ denotes the transpose, we can rewrite Eq. \ref{eq:master_eq} as:
\begin{equation}
\mathbf{P}(t+\Delta t) = \bm{\Omega}\cdot \mathbf{P}(t),
\label{eq:master_onestep}
\end{equation}
where the elements of the transition matrix $\bm{\Omega}$ are given by:
\begin{multline}
\Omega_{m,m'}=\delta_{m,m'}\left(1-\Pi_+(m)-\Pi_-(m)\right)\\+\delta_{m,m'+\Delta m}\Pi_-(m)+\delta_{m,m'-\Delta m}\Pi_+(m). \label{eq:transmat}
\end{multline}
Letting $t=n\Delta t$, we can write the formal solution to Eq. \ref{eq:master_onestep} as:
\begin{equation}
\mathbf{P}(t) = \bm{\Omega}^n\cdot \mathbf{P}(0).
\label{eq:master_soln}
\end{equation}
Numerical propagation of Eq. \ref{eq:master_soln} is straightforward, and with access to $P(m,t)$ we can compute the average magnetization as a function of time:
\begin{equation}
\langle m(t) \rangle = \sum_{m=-1}^{m_\text{f}} m P(m,t), \label{eq:avg_mag}
\end{equation}
as well as the first passage time distribution $\mathcal{P}(t)$:
\begin{equation}
\mathcal{P}(t)=-\frac{\partial \mathscr{S}(t)}{\partial t} 
\label{eq:fpt}
\end{equation}
where the survival probability $\mathscr{S}(t)$ is:
\begin{equation}
\mathscr{S}(t)=\sum_{m=-1}^{m_\text{f}} P(m,t).
\end{equation}
We compare the quantities $\langle m(t) \rangle$ and $\mathcal{P}(t)$ to their counterparts computed from MC simulations in Fig. \ref{fig:avg_mag}. In Fig. \ref{fig:fpt_dist} we plot first passage time distributions of relaxation from the formerly-metastable well to the stable well. As is evident in these figures, the dynamics of both the average magnetization and its fluctuations are captured very well by MFT.

\section{Nanoparticles}

\begin{figure}[h!]
	\centering
	\subfloat[Tagged atom at center.]{%
  		\includegraphics[width=0.49\linewidth]{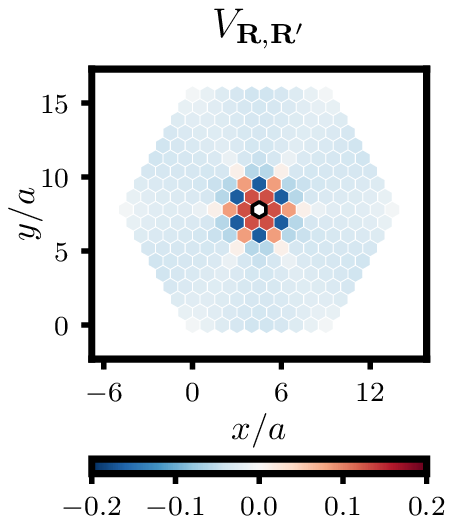}%
	}
\hfil
	\subfloat[Tagged atom on edge.]{%
  		\includegraphics[width=0.49\linewidth]{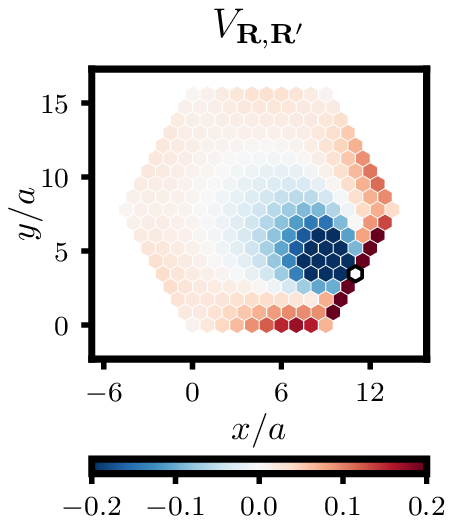}%
	}
\hfil
	\subfloat[Tagged atom between center and edge.]{%
  		\includegraphics[width=0.49\linewidth]{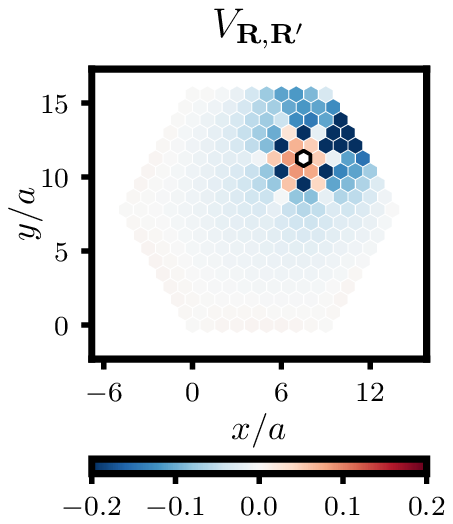}%
	}
\hfil
	\subfloat[Tagged atom at corner.]{%
  		\includegraphics[width=0.49\linewidth]{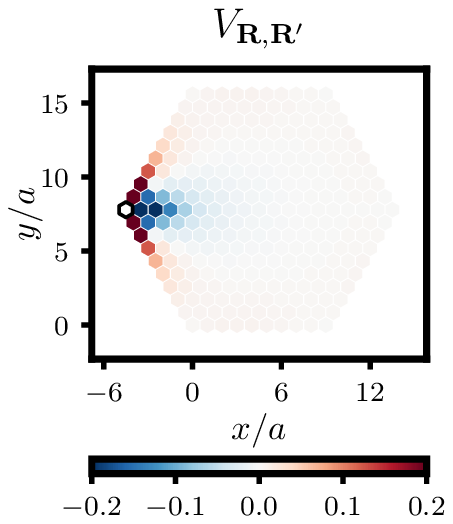}	}
\hfil
	\caption{Pair interaction function
          $V_{\mathbf{R},\mathbf{R}'}$ for different locations
          $\mathbf{R}$ of a tagged atom (outlined in black.) The value
          of $V_{\mathbf{R},\mathbf{R}'}$ for interaction with another
          atom at $\mathbf{R}'$ is indicated by color according to the
          scale shown.}
	\label{fig:nc_potential}
\end{figure}

\begin{figure}
	\centering
	\includegraphics{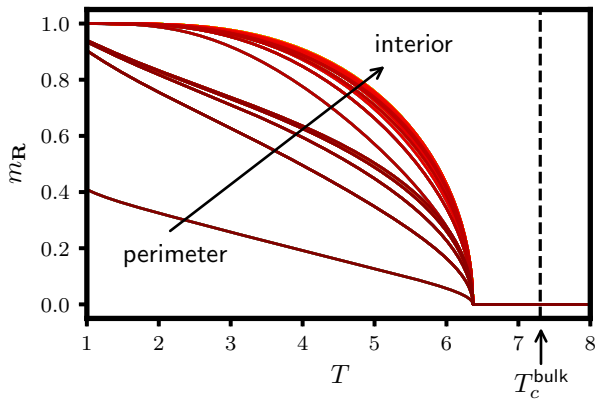}
	\caption{Numerical solutions to Eq. \ref{eq:single_site_mft}
          for the position-dependent mean-field magnetization of
          a hexagonal nanocrystal of size $N=271$ with triangular
          lattice structure. Curves with different shades of red
          represent the magnetization of different sites in the
          nanocrystal. Sites near the perimeter of the crystal have
          smaller magnetization than sites well within the interior; all
          sites transition from zero to nonzero magnetization at a
          temperature $T_c\approx 6.2$. The vertical dashed line marks
          the bulk value for $T_c$; open boundary conditions thus suppress
          the nanocrystal $T_c$ compared to its bulk value.}
	\label{fig:nc_mft_sols}
\end{figure}

\begin{figure}
	\centering
	\includegraphics{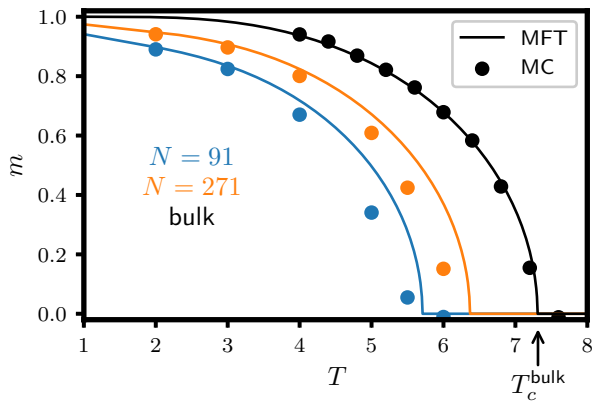}
	\caption{Nanocrystal magnetization as a function of
          temperature for different system sizes. Bulk magnetization
          versus temperature (for a system with $N=168$ subject to
          periodic boundary conditions) is included for
          comparison. Nanocrystal MFT curves were obtained as the
          numerical solutions of Eqs. \ref{eq:single_site_mft} and
          \ref{eq:net_mag}. Nanocrystal MC points were obtained as the
          minima of free energy profiles computed via umbrella
          sampling of the effective Hamiltonian,
          Eq. \ref{eq:nc_effhamil}, for each system size. Bulk MFT
          curve was computed using Eq. \ref{eq:sc_field}, and bulk MC
          points were obtained from bulk free energy minima computed
          via umbrella sampling using Eq. \ref{eq:hamil}.}
	\label{fig:nc_mc_vs_mft}
\end{figure}

\begin{figure}
	\centering
	\includegraphics{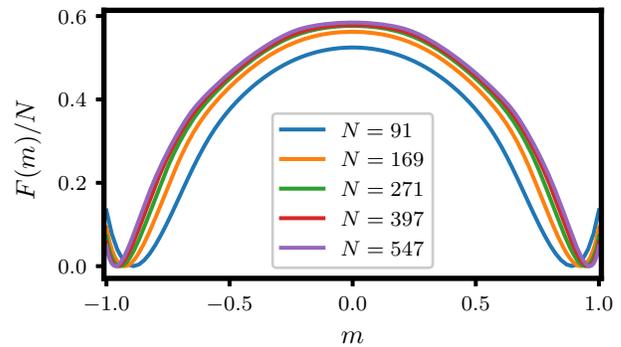}
	\caption{Nanocrystal free energies per atom for
          different system sizes at $T=3$, $h=0$. Curves were computed
          via umbrella sampling of the effective Hamiltonian,
          Eq. \ref{eq:nc_effhamil}, for each system size. These
          profiles strongly suggest a free energy barrier which
          grows linearly
          with $N$.}
	\label{fig:nc_free}
\end{figure}

In our analysis thus far we have assumed periodic boundary conditions. While this may be appropriate for a macroscopic system, many elastic materials, in particular spin-crossover compounds, have nanoscale dimensions and hence
a significant fraction of atoms at the periphery
\cite{Boldog2008,Gudyma2017,Enachescu2018,Mikolasek2019}. We therefore studied the impact of open boundary conditions on our model. While the analysis of fluctuations in $\delta a$ is insensitive to the choice of boundary conditions, broken translational invariance means that a Fourier transform will not diagonalize $\Delta \mathcal{H}$. As a result,
the required integrals in Eq. \ref{eq:effhamil_int} are
more complicated, but still numerically tractable. For a given nanocrystal size and shape, they can be performed numerically exactly to yield an effective Hamiltonian:
\begin{equation}
\mathcal{H}_{\text{eff}} = \frac{1}{2}\sum_{\mathbf{R},\mathbf{R}'}\sigma_{\mathbf{R}}V_{\mathbf{R},\mathbf{R}'}\sigma_{\mathbf{R}'} -h \sum_{\mathbf{R}}\sigma_{\mathbf{R}},
\label{eq:nc_effhamil}
\end{equation}
where due to broken translational symmetry, the effective potential depends on both $\mathbf{R}$ and $\mathbf{R}'$, not just their difference. Plots of this potential for a hexagonally-shaped nanocrystal with triangular lattice structure are shown in Fig. \ref{fig:nc_potential}.
Interactions between sites towards the interior of the crystal closely resemble bulk interactions, though interactions between sites towards the perimeter of the crystal differ significantly from bulk interactions (see \cite{Frechette2019}.) Importantly, these interior interactions largely retain the long-ranged component, meaning that MFT might still prove reasonably accurate.
Unlike in bulk, sites in the nanocrystal do not all experience the same average environment. An accurate MFT must take this spatial variation into account.
A set of self-consistent equations for the average magnetization $m_{\mathbf{R}}$
of each site in the nanocrystal can be written \cite{SM}:
\begin{equation}
m_{\mathbf{R}}=\tanh{\left(-\beta\sum_{\mathbf{R}'\neq\mathbf{R}}V_{\mathbf{R},\mathbf{R}'}m_{\mathbf{R}'}\right)}.
\label{eq:single_site_mft}
\end{equation}
An example of solutions to this set of equations, computed using the same techniques as in \cite{Frechette2019}, is shown in Fig. \ref{fig:nc_mft_sols}. The average net magnetization is then simply computed as:
\begin{equation}
m = \frac{1}{N}\sum_{\mathbf{R}}m_{\mathbf{R}}.
\label{eq:net_mag}
\end{equation}
We used Eqs. \ref{eq:single_site_mft} and \ref{eq:net_mag} to compute mean-field predictions for $m$ as a function of temperature for hexagonal nanocrystals of different sizes. Due to this finite size, thermal fluctuations will cause the system to cross the barrier separating degenerate free energy minima increasingly frequently as $T_c$ is approached from below, so that straightforward averaging of an equilibrium MC trajectory will result in $m\approx 0$. In order to compare to MFT predictions, we instead computed MC estimates for $m(T)$ by locating the minima of free energy profiles computed with umbrella sampling. MC results obtained in this way correspond quite closely with MFT predictions (see Fig. \ref{fig:nc_mc_vs_mft}), consistent with long-ranged interactions in the nanocrystal effective potential. Furthermore, we found that the height of the nanocrystal free energy barrier computed from umbrella sampling MC simulations for $T<T_c$ scales linearly with system size $N$, just as in bulk (see Fig. \ref{fig:nc_free}.) We thus anticipate similar agreement between MFT and MC for nanocrystal dynamics.

\section{Discussion}

We have thus far sidestepped a subtle, but important, caveat.
Specifically, while the factor of $1/N$ in $V^{\text{LR}}$ ensures that the energy is extensive, the arbitrarily-long interaction range means that the energy is no longer additive \cite{Mori2013, Campa2009}. In turn, this means that derivatives of the free energy $F(m)$ no longer have a definite sign, and hence its Legendre transform is no longer a single-valued function \cite{Zia2009}. In other words, the ensemble in which $m$ is fixed and the ensemble in which $m$ can fluctuate are no longer equivalent \footnote{In a related context, Vandeworpe and Newman \cite{Vandeworp1997}. previously noted inequivalence between canonical and grand canonical ensembles for a Keating model of a semiconductor mixture.}. Thus, the modulated structures observed in an ensemble with fixed magnetization -- as in our previous work \cite{Frechette2019} -- are not equilibrium states in the present ensemble, where the net magnetization can fluctuate.

Our findings have significant implications for functional elastic materials. We have shown that long-ranged interactions are a generic consequence of elastic fluctuations in lattice-mismatched solids. They should thus be operative, for instance, in spin-crossover compounds. One of the intriguing features of these compounds is the enhanced metastability of their high-spin-rich and low-spin-rich phases near room temperature \cite{Hauser1999,Hayami2001,Boldog2008,Paez-Espejo2018}, which makes them promising for use as molecular switches in next-generation data storage devices. Our results provide an underlying reason for this behavior: the extensive free energy barrier separating the two phases means that spin-crossover materials are robust to fluctuations typically responsible for the decay of metastable states. This barrier scaling also explains why transitions between high- and low-spin phases under an applied field are macroscopically sharp. We thus anticipate that our MFT will provide a simple, quantitative framework in which to explain and predict further properties of these materials.

\section*{Acknowledgments}

 This work was supported by National Science Foundation (NSF) grant CHE-1416161. This research also used resources of the National Energy Research Scientific Computing Center (NERSC), a U.S. Department of Energy Office of Science User Facility operated under Contract No. DE-AC02-05CH11231.  P.L.G. and L.B.F. acknowledge stays at the Erwin Schr{\"o}dinger Institute for Mathematics and Physics at the University of Vienna.

\nocite{Matsumoto1998}
\nocite{M.GalassiJ.DaviesJ.TheilerB.GoughG.JungmanP.AlkenM.Booth2009}
\nocite{FrenkelDSmitB2001}

\bibliography{refs_prb_master.bib}

\begin{thebibliography}{53}%
\makeatletter
\providecommand \@ifxundefined [1]{%
 \@ifx{#1\undefined}
}%
\providecommand \@ifnum [1]{%
 \ifnum #1\expandafter \@firstoftwo
 \else \expandafter \@secondoftwo
 \fi
}%
\providecommand \@ifx [1]{%
 \ifx #1\expandafter \@firstoftwo
 \else \expandafter \@secondoftwo
 \fi
}%
\providecommand \natexlab [1]{#1}%
\providecommand \enquote  [1]{``#1''}%
\providecommand \bibnamefont  [1]{#1}%
\providecommand \bibfnamefont [1]{#1}%
\providecommand \citenamefont [1]{#1}%
\providecommand \href@noop [0]{\@secondoftwo}%
\providecommand \href [0]{\begingroup \@sanitize@url \@href}%
\providecommand \@href[1]{\@@startlink{#1}\@@href}%
\providecommand \@@href[1]{\endgroup#1\@@endlink}%
\providecommand \@sanitize@url [0]{\catcode `\\12\catcode `\$12\catcode
  `\&12\catcode `\#12\catcode `\^12\catcode `\_12\catcode `\%12\relax}%
\providecommand \@@startlink[1]{}%
\providecommand \@@endlink[0]{}%
\providecommand \url  [0]{\begingroup\@sanitize@url \@url }%
\providecommand \@url [1]{\endgroup\@href {#1}{\urlprefix }}%
\providecommand \urlprefix  [0]{URL }%
\providecommand \Eprint [0]{\href }%
\providecommand \doibase [0]{https://doi.org/}%
\providecommand \selectlanguage [0]{\@gobble}%
\providecommand \bibinfo  [0]{\@secondoftwo}%
\providecommand \bibfield  [0]{\@secondoftwo}%
\providecommand \translation [1]{[#1]}%
\providecommand \BibitemOpen [0]{}%
\providecommand \bibitemStop [0]{}%
\providecommand \bibitemNoStop [0]{.\EOS\space}%
\providecommand \EOS [0]{\spacefactor3000\relax}%
\providecommand \BibitemShut  [1]{\csname bibitem#1\endcsname}%
\let\auto@bib@innerbib\@empty
%</preamble>
\bibitem [{\citenamefont {Rice}(1954)}]{Rice1954}%
  \BibitemOpen
  \bibfield  {author} {\bibinfo {author} {\bibfnamefont {O.~K.}\ \bibnamefont
  {Rice}},\ }\href {https://doi.org/10.1063/1.1740453} {\bibfield  {journal}
  {\bibinfo  {journal} {The Journal of Chemical Physics}\ }\textbf {\bibinfo
  {volume} {22}},\ \bibinfo {pages} {1535} (\bibinfo {year}
  {1954})}\BibitemShut {NoStop}%
\bibitem [{\citenamefont {Domb}(1956)}]{Domb1956}%
  \BibitemOpen
  \bibfield  {author} {\bibinfo {author} {\bibfnamefont {C.}~\bibnamefont
  {Domb}},\ }\href {https://doi.org/10.1063/1.1743060} {\bibfield  {journal}
  {\bibinfo  {journal} {The Journal of Chemical Physics}\ }\textbf {\bibinfo
  {volume} {25}},\ \bibinfo {pages} {783} (\bibinfo {year} {1956})}\BibitemShut
  {NoStop}%
\bibitem [{\citenamefont {Baker}\ and\ \citenamefont
  {Essam}(1970)}]{Baker1970}%
  \BibitemOpen
  \bibfield  {author} {\bibinfo {author} {\bibfnamefont {G.~A.}\ \bibnamefont
  {Baker}}\ and\ \bibinfo {author} {\bibfnamefont {J.~W.}\ \bibnamefont
  {Essam}},\ }\href {https://doi.org/10.1103/PhysRevLett.24.447} {\bibfield
  {journal} {\bibinfo  {journal} {Physical Review Letters}\ }\textbf {\bibinfo
  {volume} {24}},\ \bibinfo {pages} {447} (\bibinfo {year} {1970})}\BibitemShut
  {NoStop}%
\bibitem [{\citenamefont {Oitmaa}\ and\ \citenamefont
  {Barber}(1975)}]{Oitmaa1975}%
  \BibitemOpen
  \bibfield  {author} {\bibinfo {author} {\bibfnamefont {J.}~\bibnamefont
  {Oitmaa}}\ and\ \bibinfo {author} {\bibfnamefont {M.~N.}\ \bibnamefont
  {Barber}},\ }\href {https://doi.org/10.1088/0022-3719/8/21/036} {\bibfield
  {journal} {\bibinfo  {journal} {Journal of Physics C: Solid State Physics}\
  }\textbf {\bibinfo {volume} {8}},\ \bibinfo {pages} {3653} (\bibinfo {year}
  {1975})}\BibitemShut {NoStop}%
\bibitem [{\citenamefont {L{\'{e}}tard}\ \emph {et~al.}(1999)\citenamefont
  {L{\'{e}}tard}, \citenamefont {Capes}, \citenamefont {Chastanet},
  \citenamefont {Moliner}, \citenamefont {L{\'{e}}tard}, \citenamefont {Real},\
  and\ \citenamefont {Kahn}}]{Letard1999}%
  \BibitemOpen
  \bibfield  {author} {\bibinfo {author} {\bibfnamefont {J.-F.}\ \bibnamefont
  {L{\'{e}}tard}}, \bibinfo {author} {\bibfnamefont {L.}~\bibnamefont {Capes}},
  \bibinfo {author} {\bibfnamefont {G.}~\bibnamefont {Chastanet}}, \bibinfo
  {author} {\bibfnamefont {N.}~\bibnamefont {Moliner}}, \bibinfo {author}
  {\bibfnamefont {S.}~\bibnamefont {L{\'{e}}tard}}, \bibinfo {author}
  {\bibfnamefont {J.-A.}\ \bibnamefont {Real}},\ and\ \bibinfo {author}
  {\bibfnamefont {O.}~\bibnamefont {Kahn}},\ }\href
  {https://doi.org/10.1016/S0009-2614(99)01036-2} {\bibfield  {journal}
  {\bibinfo  {journal} {Chemical Physics Letters}\ }\textbf {\bibinfo {volume}
  {313}},\ \bibinfo {pages} {115} (\bibinfo {year} {1999})}\BibitemShut
  {NoStop}%
\bibitem [{\citenamefont {Hauser}\ \emph {et~al.}(1999)\citenamefont {Hauser},
  \citenamefont {Jefti{\'{c}}}, \citenamefont {Romstedt}, \citenamefont
  {Hinek},\ and\ \citenamefont {Spiering}}]{Hauser1999}%
  \BibitemOpen
  \bibfield  {author} {\bibinfo {author} {\bibfnamefont {A.}~\bibnamefont
  {Hauser}}, \bibinfo {author} {\bibfnamefont {J.}~\bibnamefont
  {Jefti{\'{c}}}}, \bibinfo {author} {\bibfnamefont {H.}~\bibnamefont
  {Romstedt}}, \bibinfo {author} {\bibfnamefont {R.}~\bibnamefont {Hinek}},\
  and\ \bibinfo {author} {\bibfnamefont {H.}~\bibnamefont {Spiering}},\ }\href
  {https://doi.org/10.1016/S0010-8545(99)00111-3} {\bibfield  {journal}
  {\bibinfo  {journal} {Coordination Chemistry Reviews}\ }\textbf {\bibinfo
  {volume} {190-192}},\ \bibinfo {pages} {471} (\bibinfo {year}
  {1999})}\BibitemShut {NoStop}%
\bibitem [{\citenamefont {Real}\ \emph {et~al.}(2005)\citenamefont {Real},
  \citenamefont {Gaspar},\ and\ \citenamefont {Mu{\~{n}}oz}}]{Real2005}%
  \BibitemOpen
  \bibfield  {author} {\bibinfo {author} {\bibfnamefont {J.~A.}\ \bibnamefont
  {Real}}, \bibinfo {author} {\bibfnamefont {A.~B.}\ \bibnamefont {Gaspar}},\
  and\ \bibinfo {author} {\bibfnamefont {M.~C.}\ \bibnamefont {Mu{\~{n}}oz}},\
  }\href {https://doi.org/10.1039/b501491c} {\bibfield  {journal} {\bibinfo
  {journal} {Dalton Transactions}\ ,\ \bibinfo {pages} {2062}} (\bibinfo {year}
  {2005})}\BibitemShut {NoStop}%
\bibitem [{\citenamefont {Gutlich}\ \emph {et~al.}(2005)\citenamefont
  {Gutlich}, \citenamefont {Ksenofontov},\ and\ \citenamefont
  {Gaspar}}]{GUTLICH2005}%
  \BibitemOpen
  \bibfield  {author} {\bibinfo {author} {\bibfnamefont {P.}~\bibnamefont
  {Gutlich}}, \bibinfo {author} {\bibfnamefont {V.}~\bibnamefont
  {Ksenofontov}},\ and\ \bibinfo {author} {\bibfnamefont {A.}~\bibnamefont
  {Gaspar}},\ }\href {https://doi.org/10.1016/j.ccr.2005.01.022} {\bibfield
  {journal} {\bibinfo  {journal} {Coordination Chemistry Reviews}\ }\textbf
  {\bibinfo {volume} {249}},\ \bibinfo {pages} {1811} (\bibinfo {year}
  {2005})}\BibitemShut {NoStop}%
\bibitem [{\citenamefont {Konishi}\ \emph {et~al.}(2008)\citenamefont
  {Konishi}, \citenamefont {Tokoro}, \citenamefont {Nishino},\ and\
  \citenamefont {Miyashita}}]{Konishi2008}%
  \BibitemOpen
  \bibfield  {author} {\bibinfo {author} {\bibfnamefont {Y.}~\bibnamefont
  {Konishi}}, \bibinfo {author} {\bibfnamefont {H.}~\bibnamefont {Tokoro}},
  \bibinfo {author} {\bibfnamefont {M.}~\bibnamefont {Nishino}},\ and\ \bibinfo
  {author} {\bibfnamefont {S.}~\bibnamefont {Miyashita}},\ }\href
  {https://doi.org/10.1103/PhysRevLett.100.067206} {\bibfield  {journal}
  {\bibinfo  {journal} {Physical Review Letters}\ }\textbf {\bibinfo {volume}
  {100}},\ \bibinfo {pages} {18} (\bibinfo {year} {2008})}\BibitemShut
  {NoStop}%
\bibitem [{\citenamefont {D{\"{u}}nweg}\ and\ \citenamefont
  {Landau}(1993)}]{Dunweg1993}%
  \BibitemOpen
  \bibfield  {author} {\bibinfo {author} {\bibfnamefont {B.}~\bibnamefont
  {D{\"{u}}nweg}}\ and\ \bibinfo {author} {\bibfnamefont {D.~P.}\ \bibnamefont
  {Landau}},\ }\href {https://doi.org/10.1103/PhysRevB.48.14182} {\bibfield
  {journal} {\bibinfo  {journal} {Physical Review B}\ }\textbf {\bibinfo
  {volume} {48}},\ \bibinfo {pages} {14182} (\bibinfo {year}
  {1993})}\BibitemShut {NoStop}%
\bibitem [{\citenamefont {Vandeworp}\ and\ \citenamefont
  {Newman}(1997)}]{Vandeworp1997}%
  \BibitemOpen
  \bibfield  {author} {\bibinfo {author} {\bibfnamefont {E.}~\bibnamefont
  {Vandeworp}}\ and\ \bibinfo {author} {\bibfnamefont {K.~E.}\ \bibnamefont
  {Newman}},\ }\href {https://doi.org/10.1103/PhysRevB.55.14222} {\bibfield
  {journal} {\bibinfo  {journal} {Physical Review B - Condensed Matter and
  Materials Physics}\ }\textbf {\bibinfo {volume} {55}},\ \bibinfo {pages}
  {14222} (\bibinfo {year} {1997})}\BibitemShut {NoStop}%
\bibitem [{\citenamefont {Miyashita}\ \emph {et~al.}(2008)\citenamefont
  {Miyashita}, \citenamefont {Konishi}, \citenamefont {Nishino}, \citenamefont
  {Tokoro},\ and\ \citenamefont {Rikvold}}]{Miyashita2008}%
  \BibitemOpen
  \bibfield  {author} {\bibinfo {author} {\bibfnamefont {S.}~\bibnamefont
  {Miyashita}}, \bibinfo {author} {\bibfnamefont {Y.}~\bibnamefont {Konishi}},
  \bibinfo {author} {\bibfnamefont {M.}~\bibnamefont {Nishino}}, \bibinfo
  {author} {\bibfnamefont {H.}~\bibnamefont {Tokoro}},\ and\ \bibinfo {author}
  {\bibfnamefont {P.~A.}\ \bibnamefont {Rikvold}},\ }\href
  {https://doi.org/10.1103/PhysRevB.77.014105} {\bibfield  {journal} {\bibinfo
  {journal} {Physical Review B}\ }\textbf {\bibinfo {volume} {77}},\ \bibinfo
  {pages} {014105} (\bibinfo {year} {2008})}\BibitemShut {NoStop}%
\bibitem [{\citenamefont {Miyashita}\ \emph {et~al.}(2009)\citenamefont
  {Miyashita}, \citenamefont {Rikvold}, \citenamefont {Mori}, \citenamefont
  {Konishi}, \citenamefont {Nishino},\ and\ \citenamefont
  {Tokoro}}]{Miyashita2009}%
  \BibitemOpen
  \bibfield  {author} {\bibinfo {author} {\bibfnamefont {S.}~\bibnamefont
  {Miyashita}}, \bibinfo {author} {\bibfnamefont {P.~A.}\ \bibnamefont
  {Rikvold}}, \bibinfo {author} {\bibfnamefont {T.}~\bibnamefont {Mori}},
  \bibinfo {author} {\bibfnamefont {Y.}~\bibnamefont {Konishi}}, \bibinfo
  {author} {\bibfnamefont {M.}~\bibnamefont {Nishino}},\ and\ \bibinfo {author}
  {\bibfnamefont {H.}~\bibnamefont {Tokoro}},\ }\href
  {https://doi.org/10.1103/PhysRevB.80.064414} {\bibfield  {journal} {\bibinfo
  {journal} {Physical Review B - Condensed Matter and Materials Physics}\
  }\textbf {\bibinfo {volume} {80}},\ \bibinfo {pages} {1} (\bibinfo {year}
  {2009})}\BibitemShut {NoStop}%
\bibitem [{\citenamefont {Frechette}\ \emph {et~al.}(2019)\citenamefont
  {Frechette}, \citenamefont {Dellago},\ and\ \citenamefont
  {Geissler}}]{Frechette2019}%
  \BibitemOpen
  \bibfield  {author} {\bibinfo {author} {\bibfnamefont {L.~B.}\ \bibnamefont
  {Frechette}}, \bibinfo {author} {\bibfnamefont {C.}~\bibnamefont {Dellago}},\
  and\ \bibinfo {author} {\bibfnamefont {P.~L.}\ \bibnamefont {Geissler}},\
  }\href {https://doi.org/10.1103/PhysRevLett.123.135701} {\bibfield  {journal}
  {\bibinfo  {journal} {Physical Review Letters}\ }\textbf {\bibinfo {volume}
  {123}},\ \bibinfo {pages} {135701} (\bibinfo {year} {2019})},\ \Eprint
  {https://arxiv.org/abs/1906.08304} {arXiv:1906.08304} \BibitemShut {NoStop}%
\bibitem [{SM()}]{SM}%
  \BibitemOpen
  \href@noop {} {}\bibinfo {note} {See Supplemental Material at [URL will be
  inserted by publisher], which includes Refs. [50-52], for
  details.}\BibitemShut {Stop}%
\bibitem [{Note1()}]{Note1}%
  \BibitemOpen
  \bibinfo {note} {Note that we have dropped the $\delta $ in front of
  $\protect \mathaccentV {tilde}07E{\sigma }_{\protect \mathbf {q}}$. That is
  because $\delta \protect \mathaccentV {tilde}07E{\sigma }_{\protect \mathbf
  {q}}=\protect \mathaccentV {tilde}07E{\sigma }_{\protect \mathbf {q}}-\delta
  _{\protect \mathbf {q},0}\protect \mathaccentV {tilde}07E{\sigma }_{0}$, but
  $\protect \mathaccentV {tilde}07E{V}_0=0$, so $\protect \mathaccentV
  {tilde}07E{\sigma }_0$ simply does not contribute to the sum.}\BibitemShut
  {Stop}%
\bibitem [{\citenamefont {Dantchev}\ and\ \citenamefont
  {Rudnick}(2001)}]{Dantchev2001}%
  \BibitemOpen
  \bibfield  {author} {\bibinfo {author} {\bibfnamefont {D.}~\bibnamefont
  {Dantchev}}\ and\ \bibinfo {author} {\bibfnamefont {J.}~\bibnamefont
  {Rudnick}},\ }\href {https://doi.org/10.1007/s100510170201} {\bibfield
  {journal} {\bibinfo  {journal} {The European Physical Journal B}\ }\textbf
  {\bibinfo {volume} {21}},\ \bibinfo {pages} {251} (\bibinfo {year}
  {2001})}\BibitemShut {NoStop}%
\bibitem [{\citenamefont {Stein}\ and\ \citenamefont
  {Shakarchi}(2003)}]{Stein2003}%
  \BibitemOpen
  \bibfield  {author} {\bibinfo {author} {\bibfnamefont {E.~M.}\ \bibnamefont
  {Stein}}\ and\ \bibinfo {author} {\bibfnamefont {R.}~\bibnamefont
  {Shakarchi}},\ }\href@noop {} {\emph {\bibinfo {title} {{Complex
  Analysis}}}},\ \bibinfo {edition} {1st}\ ed.\ (\bibinfo  {publisher}
  {Princeton University Press},\ \bibinfo {address} {Princeton, New Jersey},\
  \bibinfo {year} {2003})\BibitemShut {NoStop}%
\bibitem [{Note2()}]{Note2}%
  \BibitemOpen
  \bibinfo {note} {The function $V_{\protect \mathbf {R}}^{\protect \text
  {SR}}$ is generally anisotropic; for the triangular lattice, its slowest
  decay is $1/|\protect \mathbf {R}|^4$ along (certain linear combinations of)
  triangular lattice basis vectors \cite {Frechette2019}.}\BibitemShut {Stop}%
\bibitem [{\citenamefont {Willenbacher}\ and\ \citenamefont
  {Spiering}(1988)}]{Willenbacher1988}%
  \BibitemOpen
  \bibfield  {author} {\bibinfo {author} {\bibfnamefont {N.}~\bibnamefont
  {Willenbacher}}\ and\ \bibinfo {author} {\bibfnamefont {H.}~\bibnamefont
  {Spiering}},\ }\href {https://doi.org/10.1088/0022-3719/21/8/017} {\bibfield
  {journal} {\bibinfo  {journal} {Journal of Physics C: Solid State Physics}\
  }\textbf {\bibinfo {volume} {21}},\ \bibinfo {pages} {1423} (\bibinfo {year}
  {1988})}\BibitemShut {NoStop}%
\bibitem [{\citenamefont {Spiering}\ and\ \citenamefont
  {Willenbacher}(1989)}]{Spiering1989}%
  \BibitemOpen
  \bibfield  {author} {\bibinfo {author} {\bibfnamefont {H.}~\bibnamefont
  {Spiering}}\ and\ \bibinfo {author} {\bibfnamefont {N.}~\bibnamefont
  {Willenbacher}},\ }\href {https://doi.org/10.1088/0953-8984/1/50/011}
  {\bibfield  {journal} {\bibinfo  {journal} {Journal of Physics: Condensed
  Matter}\ }\textbf {\bibinfo {volume} {1}},\ \bibinfo {pages} {10089}
  (\bibinfo {year} {1989})}\BibitemShut {NoStop}%
\bibitem [{\citenamefont {K{\"{o}}hler}\ \emph {et~al.}(1990)\citenamefont
  {K{\"{o}}hler}, \citenamefont {Jakobi}, \citenamefont {Meissner},
  \citenamefont {Wiehl}, \citenamefont {Spiering},\ and\ \citenamefont
  {G{\"{u}}tlich}}]{Kohler1990}%
  \BibitemOpen
  \bibfield  {author} {\bibinfo {author} {\bibfnamefont {C.}~\bibnamefont
  {K{\"{o}}hler}}, \bibinfo {author} {\bibfnamefont {R.}~\bibnamefont
  {Jakobi}}, \bibinfo {author} {\bibfnamefont {E.}~\bibnamefont {Meissner}},
  \bibinfo {author} {\bibfnamefont {L.}~\bibnamefont {Wiehl}}, \bibinfo
  {author} {\bibfnamefont {H.}~\bibnamefont {Spiering}},\ and\ \bibinfo
  {author} {\bibfnamefont {P.}~\bibnamefont {G{\"{u}}tlich}},\ }\href
  {https://doi.org/10.1016/0022-3697(90)90052-H} {\bibfield  {journal}
  {\bibinfo  {journal} {Journal of Physics and Chemistry of Solids}\ }\textbf
  {\bibinfo {volume} {51}},\ \bibinfo {pages} {239} (\bibinfo {year}
  {1990})}\BibitemShut {NoStop}%
\bibitem [{\citenamefont {Boukheddaden}\ \emph
  {et~al.}(2000{\natexlab{a}})\citenamefont {Boukheddaden}, \citenamefont
  {Shteto}, \citenamefont {H{\^{o}}o},\ and\ \citenamefont
  {Varret}}]{Boukheddaden2000}%
  \BibitemOpen
  \bibfield  {author} {\bibinfo {author} {\bibfnamefont {K.}~\bibnamefont
  {Boukheddaden}}, \bibinfo {author} {\bibfnamefont {I.}~\bibnamefont
  {Shteto}}, \bibinfo {author} {\bibfnamefont {B.}~\bibnamefont {H{\^{o}}o}},\
  and\ \bibinfo {author} {\bibfnamefont {F.}~\bibnamefont {Varret}},\ }\href
  {https://doi.org/10.1103/PhysRevB.62.14796} {\bibfield  {journal} {\bibinfo
  {journal} {Physical Review B}\ }\textbf {\bibinfo {volume} {62}},\ \bibinfo
  {pages} {14796} (\bibinfo {year} {2000}{\natexlab{a}})}\BibitemShut {NoStop}%
\bibitem [{\citenamefont {Boukheddaden}\ \emph
  {et~al.}(2000{\natexlab{b}})\citenamefont {Boukheddaden}, \citenamefont
  {Shteto}, \citenamefont {H{\^{o}}o},\ and\ \citenamefont
  {Varret}}]{Boukheddaden2000a}%
  \BibitemOpen
  \bibfield  {author} {\bibinfo {author} {\bibfnamefont {K.}~\bibnamefont
  {Boukheddaden}}, \bibinfo {author} {\bibfnamefont {I.}~\bibnamefont
  {Shteto}}, \bibinfo {author} {\bibfnamefont {B.}~\bibnamefont {H{\^{o}}o}},\
  and\ \bibinfo {author} {\bibfnamefont {F.}~\bibnamefont {Varret}},\ }\href
  {https://doi.org/10.1103/PhysRevB.62.14806} {\bibfield  {journal} {\bibinfo
  {journal} {Physical Review B}\ }\textbf {\bibinfo {volume} {62}},\ \bibinfo
  {pages} {14806} (\bibinfo {year} {2000}{\natexlab{b}})}\BibitemShut {NoStop}%
\bibitem [{\citenamefont {Fourati}\ \emph {et~al.}(2018)\citenamefont
  {Fourati}, \citenamefont {Milin}, \citenamefont {Slimani}, \citenamefont
  {Chastanet}, \citenamefont {Abid}, \citenamefont {Triki},\ and\ \citenamefont
  {Boukheddaden}}]{Fourati2018}%
  \BibitemOpen
  \bibfield  {author} {\bibinfo {author} {\bibfnamefont {H.}~\bibnamefont
  {Fourati}}, \bibinfo {author} {\bibfnamefont {E.}~\bibnamefont {Milin}},
  \bibinfo {author} {\bibfnamefont {A.}~\bibnamefont {Slimani}}, \bibinfo
  {author} {\bibfnamefont {G.}~\bibnamefont {Chastanet}}, \bibinfo {author}
  {\bibfnamefont {Y.}~\bibnamefont {Abid}}, \bibinfo {author} {\bibfnamefont
  {S.}~\bibnamefont {Triki}},\ and\ \bibinfo {author} {\bibfnamefont
  {K.}~\bibnamefont {Boukheddaden}},\ }\href
  {https://doi.org/10.1039/C8CP00868J} {\bibfield  {journal} {\bibinfo
  {journal} {Physical Chemistry Chemical Physics}\ }\textbf {\bibinfo {volume}
  {20}},\ \bibinfo {pages} {10142} (\bibinfo {year} {2018})}\BibitemShut
  {NoStop}%
\bibitem [{\citenamefont {Schulz}\ \emph {et~al.}(2005)\citenamefont {Schulz},
  \citenamefont {D{\"{u}}nweg}, \citenamefont {Binder},\ and\ \citenamefont
  {M{\"{u}}ller}}]{Schulz2005}%
  \BibitemOpen
  \bibfield  {author} {\bibinfo {author} {\bibfnamefont {B.~J.}\ \bibnamefont
  {Schulz}}, \bibinfo {author} {\bibfnamefont {B.}~\bibnamefont
  {D{\"{u}}nweg}}, \bibinfo {author} {\bibfnamefont {K.}~\bibnamefont
  {Binder}},\ and\ \bibinfo {author} {\bibfnamefont {M.}~\bibnamefont
  {M{\"{u}}ller}},\ }\href {https://doi.org/10.1103/PhysRevLett.95.096101}
  {\bibfield  {journal} {\bibinfo  {journal} {Physical Review Letters}\
  }\textbf {\bibinfo {volume} {95}},\ \bibinfo {pages} {1} (\bibinfo {year}
  {2005})}\BibitemShut {NoStop}%
\bibitem [{\citenamefont {Kac}\ \emph {et~al.}(1963)\citenamefont {Kac},
  \citenamefont {Uhlenbeck},\ and\ \citenamefont {Hemmer}}]{Kac1963}%
  \BibitemOpen
  \bibfield  {author} {\bibinfo {author} {\bibfnamefont {M.}~\bibnamefont
  {Kac}}, \bibinfo {author} {\bibfnamefont {G.~E.}\ \bibnamefont {Uhlenbeck}},\
  and\ \bibinfo {author} {\bibfnamefont {P.~C.}\ \bibnamefont {Hemmer}},\
  }\href {https://doi.org/10.1063/1.1703946} {\bibfield  {journal} {\bibinfo
  {journal} {Journal of Mathematical Physics}\ }\textbf {\bibinfo {volume}
  {4}},\ \bibinfo {pages} {216} (\bibinfo {year} {1963})}\BibitemShut {NoStop}%
\bibitem [{\citenamefont {Cannas}\ \emph {et~al.}(2000)\citenamefont {Cannas},
  \citenamefont {de~Magalh{\~{a}}es},\ and\ \citenamefont
  {Tamarit}}]{Cannas2000}%
  \BibitemOpen
  \bibfield  {author} {\bibinfo {author} {\bibfnamefont {S.~A.}\ \bibnamefont
  {Cannas}}, \bibinfo {author} {\bibfnamefont {A.~C.~N.}\ \bibnamefont
  {de~Magalh{\~{a}}es}},\ and\ \bibinfo {author} {\bibfnamefont {F.~A.}\
  \bibnamefont {Tamarit}},\ }\href {https://doi.org/10.1103/PhysRevB.61.11521}
  {\bibfield  {journal} {\bibinfo  {journal} {Physical Review B}\ }\textbf
  {\bibinfo {volume} {61}},\ \bibinfo {pages} {11521} (\bibinfo {year}
  {2000})}\BibitemShut {NoStop}%
\bibitem [{\citenamefont {Vollmayr-Lee}\ and\ \citenamefont
  {Luijten}(2001)}]{Vollmayr-Lee2001}%
  \BibitemOpen
  \bibfield  {author} {\bibinfo {author} {\bibfnamefont {B.~B.}\ \bibnamefont
  {Vollmayr-Lee}}\ and\ \bibinfo {author} {\bibfnamefont {E.}~\bibnamefont
  {Luijten}},\ }\href {https://doi.org/10.1103/PhysRevE.63.031108} {\bibfield
  {journal} {\bibinfo  {journal} {Physical Review E - Statistical Physics,
  Plasmas, Fluids, and Related Interdisciplinary Topics}\ }\textbf {\bibinfo
  {volume} {63}},\ \bibinfo {pages} {1} (\bibinfo {year} {2001})}\BibitemShut
  {NoStop}%
\bibitem [{\citenamefont {Mori}(2010)}]{Mori2010}%
  \BibitemOpen
  \bibfield  {author} {\bibinfo {author} {\bibfnamefont {T.}~\bibnamefont
  {Mori}},\ }\href {https://doi.org/10.1103/PhysRevE.82.060103} {\bibfield
  {journal} {\bibinfo  {journal} {Physical Review E}\ }\textbf {\bibinfo
  {volume} {82}},\ \bibinfo {pages} {060103} (\bibinfo {year}
  {2010})}\BibitemShut {NoStop}%
\bibitem [{\citenamefont {Capel}\ \emph {et~al.}(1979)\citenamefont {Capel},
  \citenamefont {{Den Ouden}},\ and\ \citenamefont {Perk}}]{Capel1979}%
  \BibitemOpen
  \bibfield  {author} {\bibinfo {author} {\bibfnamefont {H.}~\bibnamefont
  {Capel}}, \bibinfo {author} {\bibfnamefont {L.}~\bibnamefont {{Den Ouden}}},\
  and\ \bibinfo {author} {\bibfnamefont {J.}~\bibnamefont {Perk}},\ }\href
  {https://doi.org/10.1016/0378-4371(79)90024-4} {\bibfield  {journal}
  {\bibinfo  {journal} {Physica A: Statistical Mechanics and its Applications}\
  }\textbf {\bibinfo {volume} {95}},\ \bibinfo {pages} {371} (\bibinfo {year}
  {1979})}\BibitemShut {NoStop}%
\bibitem [{\citenamefont {Nakada}\ \emph {et~al.}(2011)\citenamefont {Nakada},
  \citenamefont {Rikvold}, \citenamefont {Mori}, \citenamefont {Nishino},\ and\
  \citenamefont {Miyashita}}]{Nakada2011}%
  \BibitemOpen
  \bibfield  {author} {\bibinfo {author} {\bibfnamefont {T.}~\bibnamefont
  {Nakada}}, \bibinfo {author} {\bibfnamefont {P.~A.}\ \bibnamefont {Rikvold}},
  \bibinfo {author} {\bibfnamefont {T.}~\bibnamefont {Mori}}, \bibinfo {author}
  {\bibfnamefont {M.}~\bibnamefont {Nishino}},\ and\ \bibinfo {author}
  {\bibfnamefont {S.}~\bibnamefont {Miyashita}},\ }\href
  {https://doi.org/10.1103/PhysRevB.84.054433} {\bibfield  {journal} {\bibinfo
  {journal} {Physical Review B}\ }\textbf {\bibinfo {volume} {84}},\ \bibinfo
  {pages} {054433} (\bibinfo {year} {2011})}\BibitemShut {NoStop}%
\bibitem [{\citenamefont {Chandler}(1987)}]{Chandler1987}%
  \BibitemOpen
  \bibfield  {author} {\bibinfo {author} {\bibfnamefont {D.}~\bibnamefont
  {Chandler}},\ }\href@noop {} {\emph {\bibinfo {title} {{Introduction to
  Modern Statistical Mechanics}}}}\ (\bibinfo  {publisher} {Oxford University
  Press},\ \bibinfo {address} {New York},\ \bibinfo {year} {1987})\BibitemShut
  {NoStop}%
\bibitem [{\citenamefont {Torrie}\ and\ \citenamefont
  {Valleau}(1977)}]{Torrie1977}%
  \BibitemOpen
  \bibfield  {author} {\bibinfo {author} {\bibfnamefont {G.}~\bibnamefont
  {Torrie}}\ and\ \bibinfo {author} {\bibfnamefont {J.}~\bibnamefont
  {Valleau}},\ }\href {https://doi.org/10.1016/0021-9991(77)90121-8} {\bibfield
   {journal} {\bibinfo  {journal} {Journal of Computational Physics}\ }\textbf
  {\bibinfo {volume} {23}},\ \bibinfo {pages} {187} (\bibinfo {year}
  {1977})}\BibitemShut {NoStop}%
\bibitem [{\citenamefont {Kumar}\ \emph {et~al.}(1992)\citenamefont {Kumar},
  \citenamefont {Rosenberg}, \citenamefont {Bouzida}, \citenamefont
  {Swendsen},\ and\ \citenamefont {Kollman}}]{Kumar1992}%
  \BibitemOpen
  \bibfield  {author} {\bibinfo {author} {\bibfnamefont {S.}~\bibnamefont
  {Kumar}}, \bibinfo {author} {\bibfnamefont {J.~M.}\ \bibnamefont
  {Rosenberg}}, \bibinfo {author} {\bibfnamefont {D.}~\bibnamefont {Bouzida}},
  \bibinfo {author} {\bibfnamefont {R.~H.}\ \bibnamefont {Swendsen}},\ and\
  \bibinfo {author} {\bibfnamefont {P.~A.}\ \bibnamefont {Kollman}},\ }\href
  {https://doi.org/10.1002/jcc.540130812} {\bibfield  {journal} {\bibinfo
  {journal} {Journal of Computational Chemistry}\ }\textbf {\bibinfo {volume}
  {13}},\ \bibinfo {pages} {1011} (\bibinfo {year} {1992})}\BibitemShut
  {NoStop}%
\bibitem [{\citenamefont {Binder}(1981)}]{Binder1981a}%
  \BibitemOpen
  \bibfield  {author} {\bibinfo {author} {\bibfnamefont {K.}~\bibnamefont
  {Binder}},\ }\href {https://doi.org/10.1103/PhysRevLett.47.693} {\bibfield
  {journal} {\bibinfo  {journal} {Physical Review Letters}\ }\textbf {\bibinfo
  {volume} {47}},\ \bibinfo {pages} {693} (\bibinfo {year} {1981})}\BibitemShut
  {NoStop}%
\bibitem [{\citenamefont {Cahn}(1965)}]{Cahn1965}%
  \BibitemOpen
  \bibfield  {author} {\bibinfo {author} {\bibfnamefont {J.~W.}\ \bibnamefont
  {Cahn}},\ }\href {https://doi.org/10.1063/1.1695731} {\bibfield  {journal}
  {\bibinfo  {journal} {The Journal of Chemical Physics}\ }\textbf {\bibinfo
  {volume} {42}},\ \bibinfo {pages} {93} (\bibinfo {year} {1965})}\BibitemShut
  {NoStop}%
\bibitem [{Note3()}]{Note3}%
  \BibitemOpen
  \bibinfo {note} {Mean-field dynamics can also be accessed by directly
  simulating the MC dynamics of a mean-field Hamiltonian; see \cite {SM} for
  details.}\BibitemShut {Stop}%
\bibitem [{\citenamefont {Kawasaki}(1966)}]{Kawasaki1966}%
  \BibitemOpen
  \bibfield  {author} {\bibinfo {author} {\bibfnamefont {K.}~\bibnamefont
  {Kawasaki}},\ }\href {https://doi.org/10.1103/PhysRev.145.224} {\bibfield
  {journal} {\bibinfo  {journal} {Physical Review}\ }\textbf {\bibinfo {volume}
  {145}},\ \bibinfo {pages} {224} (\bibinfo {year} {1966})}\BibitemShut
  {NoStop}%
\bibitem [{\citenamefont {Redner}(2001)}]{Redner2001}%
  \BibitemOpen
  \bibfield  {author} {\bibinfo {author} {\bibfnamefont {S.}~\bibnamefont
  {Redner}},\ }\href@noop {} {\emph {\bibinfo {title} {{A Guide to
  First-Passage Processes}}}},\ \bibinfo {edition} {1st}\ ed.\ (\bibinfo
  {publisher} {Cambridge University Press},\ \bibinfo {address} {Cambridge,
  UK},\ \bibinfo {year} {2001})\BibitemShut {NoStop}%
\bibitem [{\citenamefont {Boldog}\ \emph {et~al.}(2008)\citenamefont {Boldog},
  \citenamefont {Gaspar}, \citenamefont {Mart{\'{i}}nez}, \citenamefont
  {Pardo-Iba{\~{n}}ez}, \citenamefont {Ksenofontov}, \citenamefont
  {Bhattacharjee}, \citenamefont {G{\"{u}}tlich},\ and\ \citenamefont
  {Real}}]{Boldog2008}%
  \BibitemOpen
  \bibfield  {author} {\bibinfo {author} {\bibfnamefont {I.}~\bibnamefont
  {Boldog}}, \bibinfo {author} {\bibfnamefont {A.~B.}\ \bibnamefont {Gaspar}},
  \bibinfo {author} {\bibfnamefont {V.}~\bibnamefont {Mart{\'{i}}nez}},
  \bibinfo {author} {\bibfnamefont {P.}~\bibnamefont {Pardo-Iba{\~{n}}ez}},
  \bibinfo {author} {\bibfnamefont {V.}~\bibnamefont {Ksenofontov}}, \bibinfo
  {author} {\bibfnamefont {A.}~\bibnamefont {Bhattacharjee}}, \bibinfo {author}
  {\bibfnamefont {P.}~\bibnamefont {G{\"{u}}tlich}},\ and\ \bibinfo {author}
  {\bibfnamefont {J.~A.}\ \bibnamefont {Real}},\ }\href
  {https://doi.org/10.1002/anie.200801673} {\bibfield  {journal} {\bibinfo
  {journal} {Angewandte Chemie International Edition}\ }\textbf {\bibinfo
  {volume} {47}},\ \bibinfo {pages} {6433} (\bibinfo {year}
  {2008})}\BibitemShut {NoStop}%
\bibitem [{\citenamefont {Gudyma}\ \emph {et~al.}(2017)\citenamefont {Gudyma},
  \citenamefont {Ivashko},\ and\ \citenamefont {Bob{\'{a}}k}}]{Gudyma2017}%
  \BibitemOpen
  \bibfield  {author} {\bibinfo {author} {\bibfnamefont {I.}~\bibnamefont
  {Gudyma}}, \bibinfo {author} {\bibfnamefont {V.}~\bibnamefont {Ivashko}},\
  and\ \bibinfo {author} {\bibfnamefont {A.}~\bibnamefont {Bob{\'{a}}k}},\
  }\href {https://doi.org/10.1186/s11671-017-1844-z} {\bibfield  {journal}
  {\bibinfo  {journal} {Nanoscale Research Letters}\ }\textbf {\bibinfo
  {volume} {12}},\ \bibinfo {pages} {101} (\bibinfo {year} {2017})}\BibitemShut
  {NoStop}%
\bibitem [{\citenamefont {Enachescu}\ and\ \citenamefont
  {Nicolazzi}(2018)}]{Enachescu2018}%
  \BibitemOpen
  \bibfield  {author} {\bibinfo {author} {\bibfnamefont {C.}~\bibnamefont
  {Enachescu}}\ and\ \bibinfo {author} {\bibfnamefont {W.}~\bibnamefont
  {Nicolazzi}},\ }\href {https://doi.org/10.1016/j.crci.2018.02.004} {\bibfield
   {journal} {\bibinfo  {journal} {Comptes Rendus Chimie}\ }\textbf {\bibinfo
  {volume} {21}},\ \bibinfo {pages} {1179} (\bibinfo {year}
  {2018})}\BibitemShut {NoStop}%
\bibitem [{\citenamefont {Mikolasek}\ \emph {et~al.}(2019)\citenamefont
  {Mikolasek}, \citenamefont {Ridier}, \citenamefont {Bessas}, \citenamefont
  {Cerantola}, \citenamefont {F{\'{e}}lix}, \citenamefont {Chaboussant},
  \citenamefont {Piedrahita-Bello}, \citenamefont {Angulo-Cervera},
  \citenamefont {Godard}, \citenamefont {Nicolazzi}, \citenamefont {Salmon},
  \citenamefont {Moln{\'{a}}r},\ and\ \citenamefont
  {Bousseksou}}]{Mikolasek2019}%
  \BibitemOpen
  \bibfield  {author} {\bibinfo {author} {\bibfnamefont {M.}~\bibnamefont
  {Mikolasek}}, \bibinfo {author} {\bibfnamefont {K.}~\bibnamefont {Ridier}},
  \bibinfo {author} {\bibfnamefont {D.}~\bibnamefont {Bessas}}, \bibinfo
  {author} {\bibfnamefont {V.}~\bibnamefont {Cerantola}}, \bibinfo {author}
  {\bibfnamefont {G.}~\bibnamefont {F{\'{e}}lix}}, \bibinfo {author}
  {\bibfnamefont {G.}~\bibnamefont {Chaboussant}}, \bibinfo {author}
  {\bibfnamefont {M.}~\bibnamefont {Piedrahita-Bello}}, \bibinfo {author}
  {\bibfnamefont {E.}~\bibnamefont {Angulo-Cervera}}, \bibinfo {author}
  {\bibfnamefont {L.}~\bibnamefont {Godard}}, \bibinfo {author} {\bibfnamefont
  {W.}~\bibnamefont {Nicolazzi}}, \bibinfo {author} {\bibfnamefont
  {L.}~\bibnamefont {Salmon}}, \bibinfo {author} {\bibfnamefont
  {G.}~\bibnamefont {Moln{\'{a}}r}},\ and\ \bibinfo {author} {\bibfnamefont
  {A.}~\bibnamefont {Bousseksou}},\ }\href
  {https://doi.org/10.1021/acs.jpclett.9b00335} {\bibfield  {journal} {\bibinfo
   {journal} {The Journal of Physical Chemistry Letters}\ }\textbf {\bibinfo
  {volume} {10}},\ \bibinfo {pages} {1511} (\bibinfo {year}
  {2019})}\BibitemShut {NoStop}%
\bibitem [{\citenamefont {Mori}(2013)}]{Mori2013}%
  \BibitemOpen
  \bibfield  {author} {\bibinfo {author} {\bibfnamefont {T.}~\bibnamefont
  {Mori}},\ }\href {https://doi.org/10.1103/PhysRevLett.111.020601} {\bibfield
  {journal} {\bibinfo  {journal} {Physical Review Letters}\ }\textbf {\bibinfo
  {volume} {111}},\ \bibinfo {pages} {020601} (\bibinfo {year}
  {2013})}\BibitemShut {NoStop}%
\bibitem [{\citenamefont {Campa}\ \emph {et~al.}(2009)\citenamefont {Campa},
  \citenamefont {Dauxois},\ and\ \citenamefont {Ruffo}}]{Campa2009}%
  \BibitemOpen
  \bibfield  {author} {\bibinfo {author} {\bibfnamefont {A.}~\bibnamefont
  {Campa}}, \bibinfo {author} {\bibfnamefont {T.}~\bibnamefont {Dauxois}},\
  and\ \bibinfo {author} {\bibfnamefont {S.}~\bibnamefont {Ruffo}},\ }\href
  {https://doi.org/10.1016/j.physrep.2009.07.001} {\bibfield  {journal}
  {\bibinfo  {journal} {Physics Reports}\ }\textbf {\bibinfo {volume} {480}},\
  \bibinfo {pages} {57} (\bibinfo {year} {2009})}\BibitemShut {NoStop}%
\bibitem [{\citenamefont {Zia}\ \emph {et~al.}(2009)\citenamefont {Zia},
  \citenamefont {Redish},\ and\ \citenamefont {McKay}}]{Zia2009}%
  \BibitemOpen
  \bibfield  {author} {\bibinfo {author} {\bibfnamefont {R.~K.~P.}\
  \bibnamefont {Zia}}, \bibinfo {author} {\bibfnamefont {E.~F.}\ \bibnamefont
  {Redish}},\ and\ \bibinfo {author} {\bibfnamefont {S.~R.}\ \bibnamefont
  {McKay}},\ }\href {https://doi.org/10.1119/1.3119512} {\bibfield  {journal}
  {\bibinfo  {journal} {American Journal of Physics}\ }\textbf {\bibinfo
  {volume} {77}},\ \bibinfo {pages} {614} (\bibinfo {year} {2009})}\BibitemShut
  {NoStop}%
\bibitem [{Note4()}]{Note4}%
  \BibitemOpen
  \bibinfo {note} {In a related context, Vandeworpe and Newman \cite
  {Vandeworp1997}. previously noted inequivalence between canonical and grand
  canonical ensembles for a Keating model of a semiconductor
  mixture.}\BibitemShut {Stop}%
\bibitem [{\citenamefont {Hayami}\ \emph {et~al.}(2001)\citenamefont {Hayami},
  \citenamefont {Gu}, \citenamefont {Yoshiki}, \citenamefont {Fujishima},\ and\
  \citenamefont {Sato}}]{Hayami2001}%
  \BibitemOpen
  \bibfield  {author} {\bibinfo {author} {\bibfnamefont {S.}~\bibnamefont
  {Hayami}}, \bibinfo {author} {\bibfnamefont {Z.-z.}\ \bibnamefont {Gu}},
  \bibinfo {author} {\bibfnamefont {H.}~\bibnamefont {Yoshiki}}, \bibinfo
  {author} {\bibfnamefont {A.}~\bibnamefont {Fujishima}},\ and\ \bibinfo
  {author} {\bibfnamefont {O.}~\bibnamefont {Sato}},\ }\href
  {https://doi.org/10.1021/ja0017920} {\bibfield  {journal} {\bibinfo
  {journal} {Journal of the American Chemical Society}\ }\textbf {\bibinfo
  {volume} {123}},\ \bibinfo {pages} {11644} (\bibinfo {year}
  {2001})}\BibitemShut {NoStop}%
\bibitem [{\citenamefont {Paez-Espejo}\ \emph {et~al.}(2018)\citenamefont
  {Paez-Espejo}, \citenamefont {Sy},\ and\ \citenamefont
  {Boukheddaden}}]{Paez-Espejo2018}%
  \BibitemOpen
  \bibfield  {author} {\bibinfo {author} {\bibfnamefont {M.}~\bibnamefont
  {Paez-Espejo}}, \bibinfo {author} {\bibfnamefont {M.}~\bibnamefont {Sy}},\
  and\ \bibinfo {author} {\bibfnamefont {K.}~\bibnamefont {Boukheddaden}},\
  }\href {https://doi.org/10.1021/jacs.8b04802} {\bibfield  {journal} {\bibinfo
   {journal} {Journal of the American Chemical Society}\ }\textbf {\bibinfo
  {volume} {140}},\ \bibinfo {pages} {11954} (\bibinfo {year}
  {2018})}\BibitemShut {NoStop}%
\bibitem [{\citenamefont {Matsumoto}\ and\ \citenamefont
  {Nishimura}(1998)}]{Matsumoto1998}%
  \BibitemOpen
  \bibfield  {author} {\bibinfo {author} {\bibfnamefont {M.}~\bibnamefont
  {Matsumoto}}\ and\ \bibinfo {author} {\bibfnamefont {T.}~\bibnamefont
  {Nishimura}},\ }\href {https://doi.org/10.1145/272991.272995} {\bibfield
  {journal} {\bibinfo  {journal} {ACM Transactions on Modeling and Computer
  Simulation}\ }\textbf {\bibinfo {volume} {8}},\ \bibinfo {pages} {3}
  (\bibinfo {year} {1998})}\BibitemShut {NoStop}%
\bibitem [{\citenamefont {{M. Galassi, J. Davies, J. Theiler, B. Gough, G.
  Jungman, P. Alken, M.
  Booth}}(2009)}]{M.GalassiJ.DaviesJ.TheilerB.GoughG.JungmanP.AlkenM.Booth2009}%
  \BibitemOpen
  \bibfield  {author} {\bibinfo {author} {\bibfnamefont {F.~R.}\ \bibnamefont
  {{M. Galassi, J. Davies, J. Theiler, B. Gough, G. Jungman, P. Alken, M.
  Booth}}},\ }\href {http://www.gnu.org/software/gsl/} {\emph {\bibinfo {title}
  {{GNU Scientific Library Reference Manual}}}},\ \bibinfo {edition} {3rd}\
  ed.\ (\bibinfo  {publisher} {Network Theory Ltd.},\ \bibinfo {year}
  {2009})\BibitemShut {NoStop}%
\bibitem [{\citenamefont {{D. Frenkel and B. Smit}}(2001)}]{FrenkelDSmitB2001}%
  \BibitemOpen
  \bibfield  {author} {\bibinfo {author} {\bibnamefont {{D. Frenkel and B.
  Smit}}},\ }\href@noop {} {\emph {\bibinfo {title} {{Understanding molecular
  simulation: from algorithms to applications}}}},\ \bibinfo {edition} {2nd}\
  ed.\ (\bibinfo  {publisher} {Academic Press},\ \bibinfo {address} {San
  Diego},\ \bibinfo {year} {2001})\ Chap.~\bibinfo {chapter} {7}\BibitemShut
  {NoStop}%
\end{thebibliography}%

\ifarXiv


\section*{Supplementary material}

# Supplemental Material for "The Origin of Mean-Field Behavior in an Elastic Ising Model"

Layne B. Frechette,[1,2] Christoph Dellago,[2,3] and Phillip L. Geissler[1,2]

[1]*Department of Chemistry, University of California, Berkeley, California 94720, USA*
[2]*Erwin Schrödinger Institute for Mathematics and Physics,*
*University of Vienna, Boltzmanngasse 9, Wien 1090, Austria*
[3]*Faculty of Physics, University of Vienna, Boltzmanngasse 5, Wien 1090, Austria*

(Dated: April 17, 2020)

## PRESSURE IN THE SMALL-MISMATCH LIMIT

In the small-mismatch limit, we can simplify the pressure-volume contribution to our Hamiltonian, which we label here as $E(pV)$:

$$E(pV) = pcNa^d. \tag{1}$$

To begin, we use the the fact that $a = \delta a + l_{AB}$:

$$E(pV) = pcN(\delta a + l_{AB})^d \tag{2}$$

$$= pcNl_{AB}^d + pcNdl_{AB}^{d-1}\delta a + \mathcal{O}(\Delta^2), \tag{3}$$

since $\delta a$ is of order $\Delta$. Dropping the constant first term and the terms of order $\mathcal{O}(\Delta^2)$, which vanish in the small-mismatch limit, we arrive at:

$$E(pV) \approx pcNdl_{AB}^{d-1}\delta a. \tag{4}$$

## MC SIMULATION DETAILS

We used MC simulations to compute both equilibrium and dynamic properties of our elastic model. For simulations of bulk crystals on a triangular lattice consisting of $N = N_x N_y$ atoms, periodic boundary conditions were imposed in the $x$ and $y$ directions. Simulations of nanocrystals employed hexagonal systems with open boundary conditions. For simulations employing the full elastic Hamiltonian $\mathcal{H}$, we used a value of $\Delta = 0.1$ for the lattice mismatch and performed simulations in the isothermal-isobaric (NPT) ensemble. This ensemble was sampled using two basic MC moves: spin flips and displacement moves. In both cases, an atom at lattice site $\mathbf{R}$ was selected at random. For spin flips, an attempt was then made to change its spin, $\sigma_{\mathbf{R}} \to -\sigma_{\mathbf{R}}$. For displacement moves, an attempt was made to change the atom's position, $\mathbf{r}_{\mathbf{R}} \to \mathbf{r}_{\mathbf{R}} + \mathbf{d}$, where $\mathbf{d} = (d_x, d_y)$ is a two dimensional vector whose components were selected uniformly at random from an interval $[-0.1, 0.1]$. Random numbers were generated using the Mersenne Twister algorithm [1] as implemented in the GNU Scientific Library (GSL) [2]. A Metropolis criterion was used to accept or reject proposed moves, ensuring detailed balance [3]:

$$P(C \to C') = \min\left[1, e^{-\beta(\mathcal{H}(C') - \mathcal{H}(C))}\right], \tag{5}$$

where $C$ and $C'$ represent configurations $\{\sigma_{\mathbf{R}}\}, \{\mathbf{r}_{\mathbf{R}}\}$ before and after the proposed move, respectively. Equilibrium simulation runs consisted of performing a large number (usually $10^6$) of MC sweeps. A single MC sweep consisted of $N$ attempted spin flips and $N$ attempted displacement moves. Constant pressure was maintained using a standard algorithm in which attempts to change the system's volume were proposed and then accepted or rejected according to a Metropolis criterion [3]. Proposed volume moves consisted of changing the total volume by an amount $\delta V$, selected uniformly at random from the interval $[-\delta V_{\max}, \delta V_{\max}]$. We chose $\delta V_{\max} = 0.01 V_{\text{init}}$, where $V_{\text{init}}$ is the volume at the beginning of a simulation run. Such volume moves were performed once every MC sweep. Initial configurations consisted of atoms arranged on a perfect triangular lattice, with a random distribution of spins, and with a volume consistent with the net magnetization. Before obtaining statistics, we equilibrated the system by running at least 100 (usually more) MC sweeps without collecting any data. Data was then recorded for different observables once every sweep.

Umbrella sampling simulations [4], used in combination with WHAM [5] to compute the equilibrium free energy as a function of magnetization, employed $n_{\text{window}} = 50$ evenly spaced harmonic biases with spring constants of strength $0.1\epsilon$.

Equilibrium simulations employing the effective Hamiltonian used the same protocols as for the full elastic Hamiltonian, except that no position moves or volume moves were necessary, and simulations were performed in the canonical ensemble.

Dynamical properties were computed via simulations of the effective Hamiltonian, employing the spin-flip moves and Metropolis criterion described previously.

Properties of the nanocrystals were computed using simulations of the nanocrystal effective Hamiltonian in the canonical ensemble. Spin-flip moves and the Metropolis criterion were used as in bulk simulations. Umbrella sampling simulations for the nanocrystals used spring constants of strength $0.2\epsilon$ for the 50 evenly spaced harmonic biases.

## CRITICAL TEMPERATURE AND SCALING ON DIFFERENT LATTICES

Here we assess mean-field predictions for the critical temperature ($T_c$) and exponents of spontaneous symmetry breaking of the elastic Ising model on different lattices. Mean-field scaling as measured by critical exponents should hold for each lattice (as long as there is a gap in the effective potential at $\mathbf{q} = 0$,) since mean-field exponents are known to be robust to the addition of short-ranged interactions [6, 7]. However, since it is a non-universal quantity, the accuracy of mean-field predictions for $T_c$ will depend on the lattice structure, in particular the magnitude of short-ranged interactions compared to the long-ranged interaction. We examine five different lattices, spanning one to three spatial dimensions, and show that each exhibits mean-field critical exponents, as expected. Mean field theory (MFT) predicts $T_c$ with reasonable accuracy for all lattices except the one-dimensional (1d) lattice. This is likely a consequence of the significant magnitude of short-ranged interactions for that lattice.

### A. Theory & Methods

As argued in the main text, MFT predicts that the critical temperature for spontaneous symmetry breaking of the magnetization $m$ is given by:

$$T_c = 2\bar{V}, \tag{6}$$

$$\bar{V} = -\sum_{\mathbf{R} \neq 0} V_{\mathbf{R}}/2. \tag{7}$$

The mean-field potential $\bar{V}$ is composed of a long-ranged part $\bar{V}^{\text{LR}} = \lim_{\mathbf{q} \to 0} \tilde{V}_{\mathbf{q}}/2$ and a short-ranged part $\bar{V}^{\text{SR}} = \bar{V} - \bar{V}^{\text{LR}}$. If the magnitude of $\bar{V}^{\text{SR}}$ is small compared to that of $\bar{V}^{\text{LR}}$ then we expect MFT to give a reasonable estimate for $T_c$. MFT also predicts [8] that the average squared magnetization $\langle m^2 \rangle$ obeys:

$$\langle m^2 \rangle = N^{-1/2} f_1(tN^{1/2}), \tag{8}$$

where $t = (T - T_c)/T_c$ and $N$ is the system size, and the Binder cumulant $U = 1 - \langle m^4 \rangle/(3\langle m^2 \rangle^2)$ obeys:

$$U = f_2(tN^{1/2}). \tag{9}$$

The scaling functions $f_1$ and $f_2$ depend only on the scaled temperature $tN^{1/2}$, and hence plots of $N^{1/2}\langle m^2 \rangle$ and $U$ versus $tN^{1/2}$ for different system sizes should fall on top of one another. Additionally, in MFT the Binder cumulant attains a universal value of $\approx 0.27$ at $T_c$ [8].

To verify these relationships, we employ MC simulations using the effective Hamiltonian for each lattice, which improves sampling since we do not have to explicitly evolve mechanical degrees of freedom. With these simulations, we compute the quantities $U$ and $\langle m^2 \rangle$ as functions of temperature for several system sizes. At each temperature and system size we equilibrate the system for $10^3$ MC sweeps (1 sweep is $N$ attempts to flip a spin) and then collect data every sweep for $10^6$ sweeps. The MC estimate for $T_c$ is given by the intersection of Binder cumulants for different system sizes.

## B. 1d Lattice

MFT does not accurately predict $T_c$ for the 1d lattice (with both nearest- and next-nearest-neighbor springs):

$$T_c^{\mathrm{MF}} \approx 3.57 \tag{10}$$

$$T_c^{\mathrm{MC}} \approx 2.79. \tag{11}$$

See Fig. 1. That is because $\bar{V}^{\mathrm{LR}} = 0.8$ is comparable to $\bar{V}^{\mathrm{SR}} = 0.987$ ($\bar{V}^{\mathrm{SR}}$ is $\approx 55\%$ of $\bar{V}$.) When $U$ and $\langle m^2 \rangle$ are scaled using $T_c^{\mathrm{MF}}$, data collapse is poor; however, when they are scaled using $T_c^{\mathrm{MC}}$, data collapse is excellent (see Figs. 2 and 3.) Thus the critical exponents of the 1d lattice have mean-field scaling, as expected.

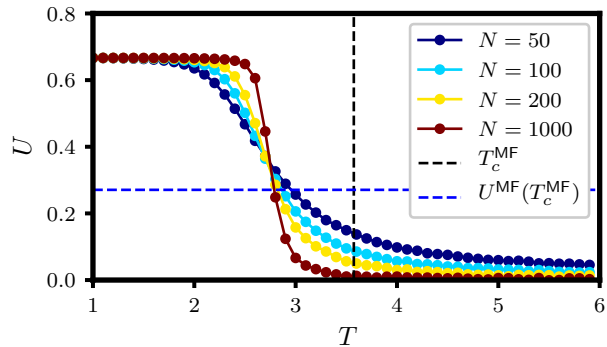

FIG. 1: Binder cumulants for the 1d lattice.

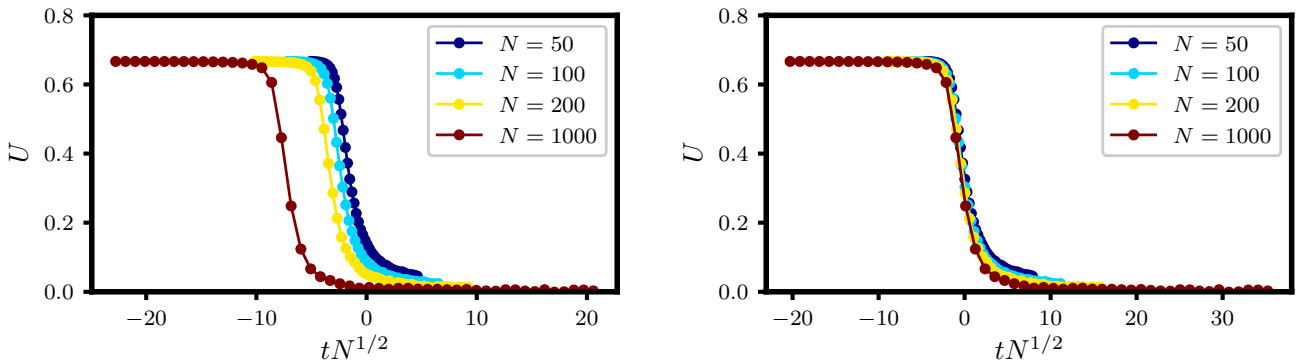

(a) $T_c$ computed from MFT.

(b) $T_c$ computed from MC.

FIG. 2: Scaling of $U$ for different choices of $T_c$ on the 1d lattice.

## C. Triangular Lattice

MFT accurately predicts $T_c$ for the 2d triangular lattice:

$$T_c^{\mathrm{MF}} \approx 7.31 \tag{12}$$

$$T_c^{\mathrm{MC}} \approx 7.20. \tag{13}$$

See Fig. 4. That is because $\bar{V}^{\mathrm{LR}} = 4.0$ is much larger in magnitude than $\bar{V}^{\mathrm{SR}} = -0.346$ ($\bar{V}^{\mathrm{SR}}$ is $\approx 8.7\%$ of $\bar{V}$.) When $U$ and $\langle m^2 \rangle$ are scaled using $T_c^{\mathrm{MF}}$, data collapse is excellent (see Figs. 4 and 5.)

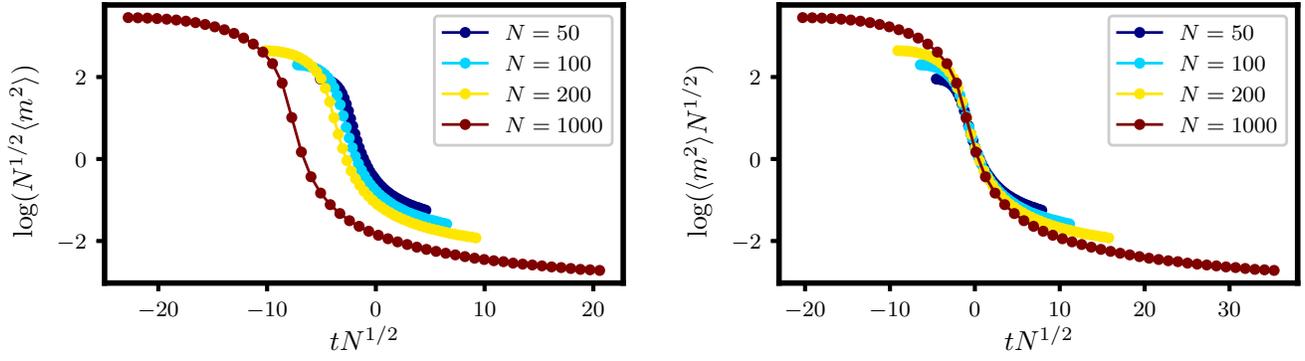

(a) $T_c$ computed from MFT.

(b) $T_c$ computed from MC.

FIG. 3: Scaling of $\langle m^2 \rangle$ for different choices of $T_c$ on the 1d lattice.

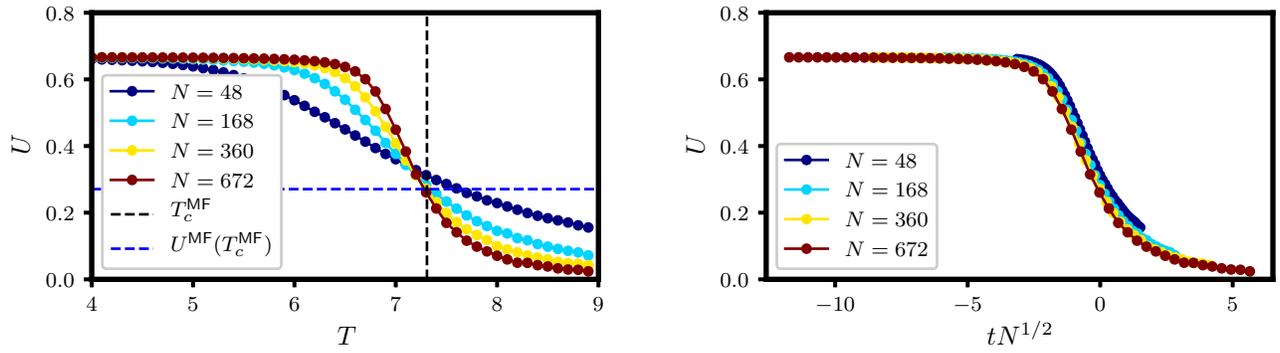

(a) Binder cumulants for the triangular lattice.

(b) Scaling of $U$ for the triangular lattice.

FIG. 4

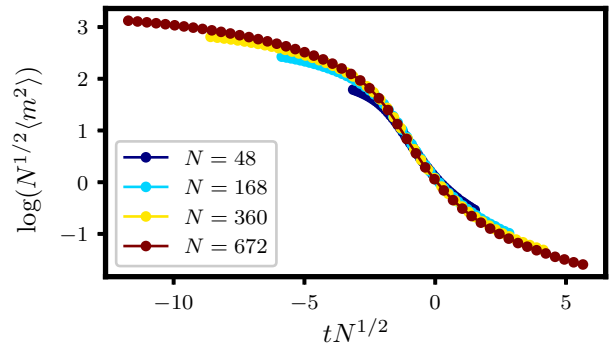

FIG. 5: Scaling of $\langle m^2 \rangle$ for the triangular lattice.

### D. Square Lattice

MFT predicts $T_c$ for the 2d square lattice (with nearest- and next-nearest-neighbor springs) with reasonable accuracy:

$$T_c^{\mathrm{MF}} \approx 11.1 \tag{14}$$

$$T_c^{\mathrm{MC}} \approx 10.6. \tag{15}$$

See Fig. 6. That is because $\bar{V}^{\mathrm{LR}} = 4.34$ is fairly large in magnitude compared to $\bar{V}^{\mathrm{SR}} = 1.20$ ($\bar{V}^{\mathrm{SR}}$ is $\approx 21.6\%$ of $\bar{V}$.) When $U$ and $\langle m^2 \rangle$ are scaled using $T_c^{\mathrm{MF}}$, data collapse is good (see Figs. 6 and 7.)

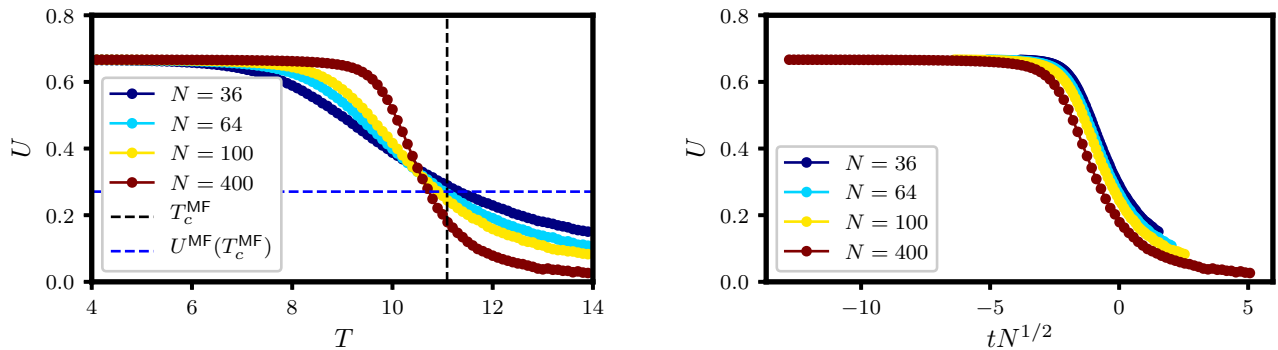

(a) Binder cumulants for the square lattice.

(b) Scaling of $U$ for the square lattice.

FIG. 6

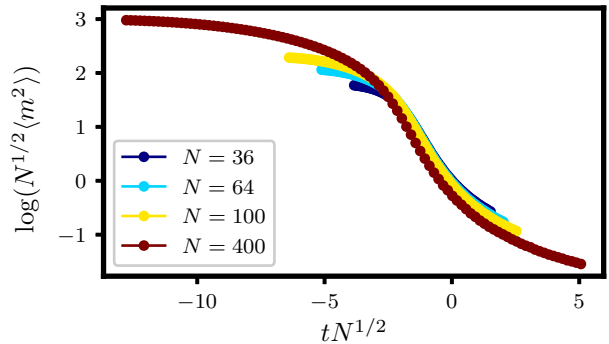

FIG. 7: Scaling of $\langle m^2 \rangle$ for the square lattice.

### E.  FCC Lattice

MFT accurately predicts $T_c$ for the 3d FCC lattice:

$$T_c^{\mathrm{MF}} \approx 18.3 \tag{16}$$

$$T_c^{\mathrm{MC}} \approx 18.0. \tag{17}$$

See Fig. 8. That is because $\bar{V}^{\mathrm{LR}} = 8$ is much larger in magnitude than $\bar{V}^{\mathrm{SR}} = 1.13$ ($\bar{V}^{\mathrm{SR}}$ is $\approx 14.1\%$ of $\bar{V}$.) When $U$ and $\langle m^2 \rangle$ are scaled using $T_c^{\mathrm{MF}}$, data collapse is excellent (see Figs. 8 and 9.)

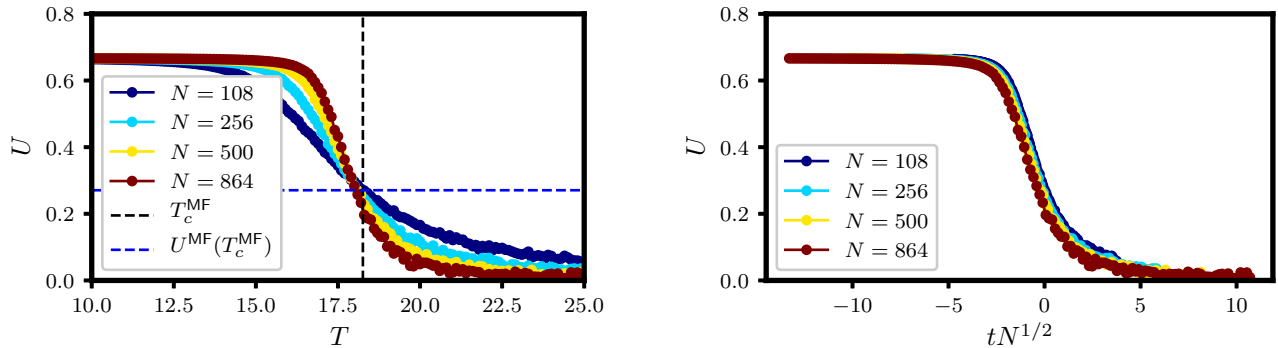

(a) Binder cumulants for the FCC lattice.

(b) Scaling of $U$ for the FCC lattice.

FIG. 8

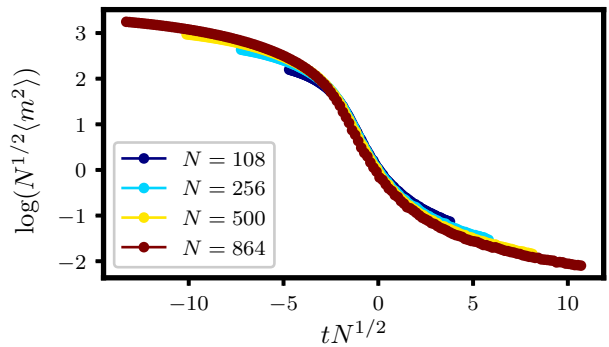

FIG. 9: Scaling of $\langle m^2 \rangle$ for the FCC lattice.

## F.   Simple Cubic Lattice

MFT accurately predicts $T_c$ for the 3d simple cubic lattice (with nearest- and next-nearest-neighbor springs):

$$T_c^{\mathrm{MF}} \approx 30.3 \tag{18}$$

$$T_c^{\mathrm{MC}} \approx 29.8. \tag{19}$$

See Fig. 10. That is because $\bar{V}^{\mathrm{LR}} = 16.5$ is much larger in magnitude than $\bar{V}^{\mathrm{SR}} = -1.31$ ($\bar{V}^{\mathrm{SR}}$ is $\approx 7.9\%$ of $\bar{V}$.) When $U$ and $\langle m^2 \rangle$ are scaled using $T_c^{\mathrm{MF}}$, data collapse is excellent (see Figs. 10 and 11.)

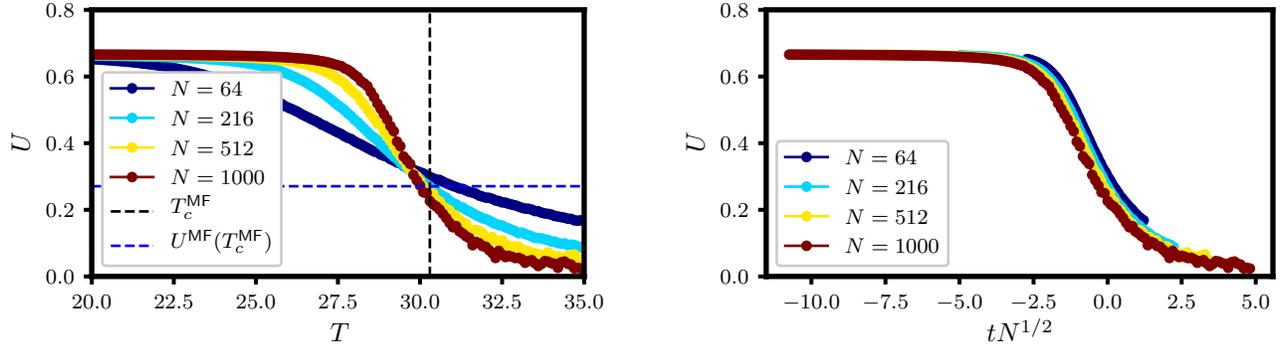

(a) Binder cumulants for the simple cubic lattice.

(b) Scaling of $U$ for the simple cubic lattice.

FIG. 10

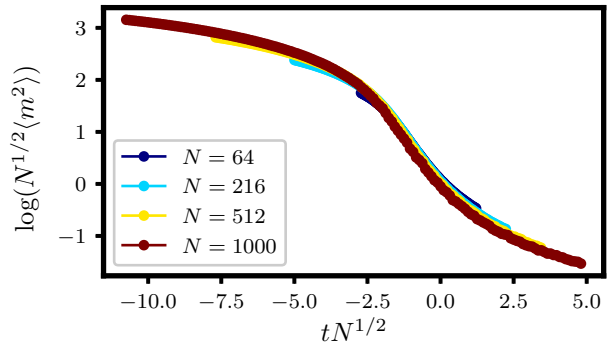

FIG. 11: Scaling of $\langle m^2 \rangle$ for the simple cubic lattice.

## MEAN-FIELD DYNAMICS

In the main text, we studied mean-field relaxation dynamics by by solving a mean-field master equation for the time-dependent magnetization probability distribution, $P(m, t)$. Alternatively, one could simply harvest many MC trajectories of mean-field Hamiltonian:

$$\mathcal{H}_{\mathrm{MF}} = -\frac{\bar{V}}{N} \sum_{\mathbf{R}, \mathbf{R}'} \sigma_{\mathbf{R}} \sigma_{\mathbf{R}'} - h \sum_{\mathbf{R}} \sigma_{\mathbf{R}}, \tag{20}$$

We label this approach "MFMC." Trajectories were initialized at configurations sampled from an equilibrium distribution $\propto \exp\left(-\beta \mathcal{H}_{\mathrm{MF}}\right)$ at fixed magnetization $m = -0.7$ and inverse temperature $\beta = 1/6$, with system size $N = 168$. Allowing the magnetization to then fluctuate and imposing an external field $h = 0.5$, each trajectory consisted of 100 MC sweeps. We harvested $10^5$ such trajectories. We compare the MFMC and master equation approaches to dynamics in Figs. 12 and 13. Excellent agreement is evident for both the average magnetization versus time and the mean first passage time distribution.

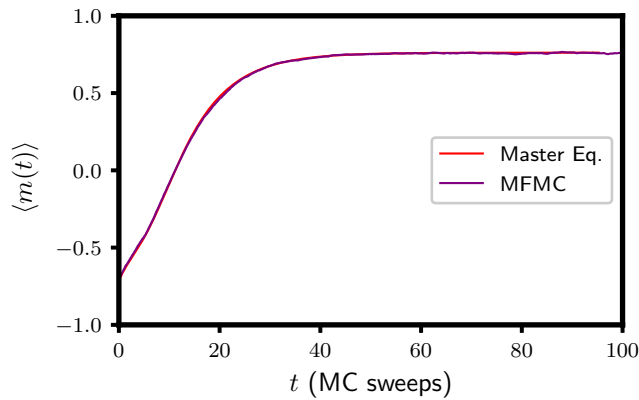

FIG. 12: Average magnetization versus time. MFMC results were averaged over $10^5$ Monte Carlo trajectories.

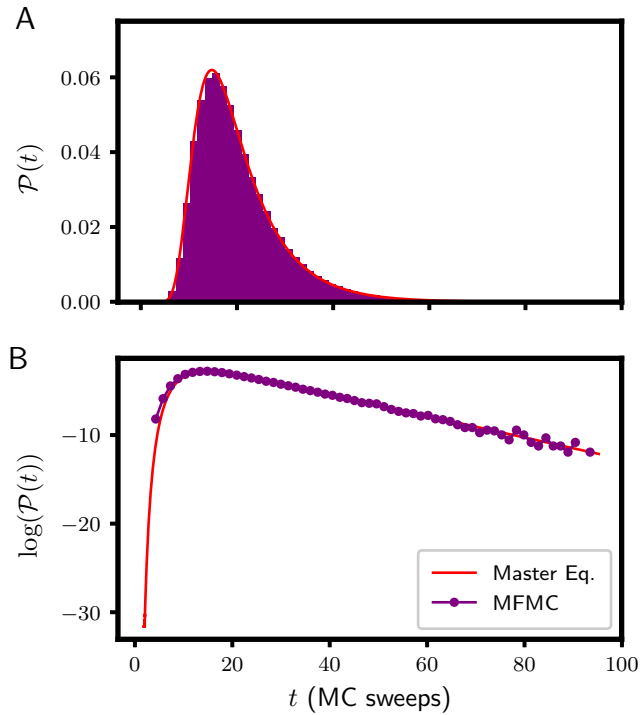

FIG. 13: First passage time distributions. **A** shows distributions on a linear scale while **B** shows distributions on a logarithmic scale. MFMC histogram was made from $10^5$ Monte Carlo trajectories.

## SINGLE-SITE MFT

Standard MFT assumes that the order parameter $m$ is spatially uniform. This assumption is a poor one in systems such as nanocrystals which are intrinsically spatially non-uniform. It is more appropriate to consider a spatially-resolved order parameter $m(\mathbf{R})$ which is capable of capturing spatial inhomogeneities. With this in mind, consider a spin Hamiltonian with pair interactions:

$$\mathcal{H} = \sum_{\mathbf{R}, \mathbf{R}' \neq \mathbf{R}} \frac{1}{2} \sigma_{\mathbf{R}} V_{\mathbf{R}, \mathbf{R}'} \sigma_{\mathbf{R}'}, \tag{21}$$

where we have excluded self-interactions. As our mean-field *ansatz* we take a one-body Hamiltonian with spatially-varying field:

$$\mathcal{H}_0 = -\sum_{\mathbf{R}} h_{\mathbf{R}} \sigma_{\mathbf{R}}. \tag{22}$$

The associated partition function is:

$$
\begin{aligned}
Q_0 &= \sum_{\{\sigma_{\mathbf{R}}\}} \exp(-\beta \mathcal{H}_0) \\
&= \sum_{\sigma_1} \exp\left(\beta h_1 \sigma_1\right) \sum_{\sigma_2} \exp\left(\beta h_2 \sigma_2\right) \cdots \\
&= \prod_{\mathbf{R}} 2 \cosh \beta h_{\mathbf{R}}.
\end{aligned}
\tag{23}
$$

The spatially-varying magnetization is easily extracted:

$$
\begin{aligned}
m(\mathbf{R}) = \langle \sigma_{\mathbf{R}} \rangle_0 &= \frac{\partial \log Q_0}{\partial \beta h_{\mathbf{R}}} \\
&= \tanh \beta h_{\mathbf{R}}.
\end{aligned}
$$

To obtain the optimal value of $h_{\mathbf{R}}$, we apply the Gibbs-Bogoliubov-Feynman bound [9]:

$$Q_{\text{est}} = Q_0 e^{-\beta\langle\Delta\mathcal{H}\rangle_0},$$

(24)

where $\Delta\mathcal{H} = \mathcal{H} - \mathcal{H}_0$. We variationally optimize $h_{\mathbf{R}}$ by setting the derivative of (the logarithm of) this partition function to zero:

$$
\begin{aligned}
0 &= \frac{\partial \log Q_{\text{est}}}{\partial \beta h_{\mathbf{R}}} \\
&= \frac{\partial \log Q_0}{\partial \beta h_{\mathbf{R}}} - \frac{\partial}{\partial \beta h_{\mathbf{R}}} \beta \langle\Delta\mathcal{H}\rangle_0
\end{aligned}
$$

The necessary average is given by:

$$\langle\Delta\mathcal{H}\rangle_0 = \frac{1}{2} \sum_{\mathbf{R},\mathbf{R}'} V_{\mathbf{R},\mathbf{R}'} \langle\sigma_{\mathbf{R}}\rangle_0 \langle\sigma_{\mathbf{R}'}\rangle_0 + \sum_{\mathbf{R}} h_{\mathbf{R}} \langle\sigma_{\mathbf{R}}\rangle_0.$$

(25)

Evaluating the derivative,

$$\frac{\partial}{\partial h_{\mathbf{R}}} \langle\Delta\mathcal{H}\rangle_0 = \left( \sum_{\mathbf{R}'} V_{\mathbf{R},\mathbf{R}'} \langle\sigma_{\mathbf{R}'}\rangle_0 + h_{\mathbf{R}} \right) \frac{\partial\langle\sigma_{\mathbf{R}}\rangle_0}{\partial h_{\mathbf{R}}} + \langle\sigma_{\mathbf{R}}\rangle_0.$$

(26)

Using the fact that $\partial \log Q_0 / \partial \beta h_{\mathbf{R}} = \langle\sigma_{\mathbf{R}}\rangle_0$, we find an expression for the optimal field:

$$h_{\mathbf{R}} = -\sum_{\mathbf{R}'} V_{\mathbf{R},\mathbf{R}'} \langle\sigma_{\mathbf{R}'}\rangle_0.$$

(27)

This finally yields a self-consistent equation for the spatially-varying order parameter:

$$m(\mathbf{R}) = \tanh\left( -\beta \sum_{\mathbf{R}'} V_{\mathbf{R},\mathbf{R}'} m(\mathbf{R}') \right).$$

(28)

[1] M. Matsumoto and T. Nishimura, ACM Transactions on Modeling and Computer Simulation **8**, 3 (1998).

[2] F. R. M. Galassi, J. Davies, J. Theiler, B. Gough, G. Jungman, P. Alken, M. Booth, *GNU Scientific Library Reference Manual*, 3rd ed. (Network Theory Ltd., 2009).

[3] D. Frenkel and B. Smit, *Understanding molecular simulation: from algorithms to applications*, 2nd ed. (Academic Press, San Diego, 2001) Chap. 7.

[4] G. Torrie and J. Valleau, Journal of Computational Physics **23**, 187 (1977).

[5] S. Kumar, J. M. Rosenberg, D. Bouzida, R. H. Swendsen, and P. A. Kollman, Journal of Computational Chemistry **13**, 1011 (1992).

[6] H. Capel, L. Den Ouden, and J. Perk, Physica A: Statistical Mechanics and its Applications **95**, 371 (1979).

[7] T. Nakada, P. A. Rikvold, T. Mori, M. Nishino, and S. Miyashita, Physical Review B **84**, 054433 (2011).

[8] S. Miyashita, Y. Konishi, M. Nishino, H. Tokoro, and P. A. Rikvold, Physical Review B **77**, 014105 (2008).

[9] D. Chandler, *Introduction to Modern Statistical Mechanics* (Oxford University Press, New York, 1987).


\fi

\end{document}